# Optical wood with switchable solar transmittance for all-round thermal management

*He Gao[#ab], Ying Li[#c], Yanjun Xie[a], Daxin Liang[*a], Jian Li[b], Yonggui Wang[a], Zefang Xiao[b], Haigang Wang[b], Wentao Gan[*b], Lorenzo Pattelli[*d,e], and Hongbo Xu[*f].*

e-mail: daxin.liang@nefu.edu.cn, iamxhb@hit.edu.cn, wtgan@nefu.edu.cn, pattelli@lens.unifi.it

[a] Key Laboratory of Bio-based Material Science & Technology (Northeast Forestry University), Ministry of Education, Hexing Road 26, Harbin, 150040, P.R. China

[b] Engineering Research Center of Advanced Wooden Materials (Northeast Forestry University), Ministry of Education, Hexing Road 26, Harbin, 150040, P.R. China

[c] State Key Laboratory of Extreme Photonics and Instrumentation, ZJU-Hangzhou Global Scientific and Technological Innovation Center, Zhejiang University, Hangzhou, 310027, 310027, P.R. China

[d] Istituto Nazionale di Ricerca Metrologica (INRiM), Turin, 10135, Italy

[e] European Laboratory for Non-linear Spectroscopy (LENS), Sesto Fiorentino, 50019, Italy

[f] MIIT Key Laboratory of Critical Materials Technology for New Energy Conversion and Storage, School of Chemistry and Chemical Engineering, Harbin Institute of Technology, Harbin, 150001, P.R. China.

# These authors contributed equally to this work.

**Abstract** Technologies enabling passive daytime radiative cooling and daylight harvesting are highly relevant for energy-efficient buildings. Despite recent progress demonstrated with passively cooling polymer coatings, however, it remains challenging to combine also a passive heat gain mechanism into a single substrate for all-round thermal management. Herein, we developed an optical wood (OW) with switchable transmittance of solar irradiation enabled by the hierarchically porous structure, ultralow absorption in solar spectrum and high infrared absorption of cellulose nanofibers. After delignification, the OW shows a high solar reflectance (94.9%) in the visible and high broadband emissivity (0.93) in the infrared region (2.5–25 μm). Owing to the exceptional mass transport of its aligned cellulose nanofibers, OW can quickly switch to a new highly transparent state following phenylethanol impregnation. The solar transmittance of optical wood (OW-II state) can reach 68.4% from 250 to 2500 nm. The switchable OW exhibits efficient radiative cooling to 4.5 °C below ambient temperature in summer (cooling power 81.4 W m$^{-2}$), and daylight heating to 5.6 °C above the temperature of natural wood in winter (heating power 229.5 W m$^{-2}$), suggesting its promising role as a low-cost and sustainable solution to all-season thermal management applications.

*Keywords:* Optical wood, Radiative cooling, Daylight harvesting, All-season thermal management, Switchable.

# 1. Introduction

Thermoregulation in building represents a major energy need in our society. In the EU, buildings alone are responsible for 40% of our energy consumption and 36% of greenhouse gas emissions, representing the single largest energy consumer in Europe [1]. Data released by the United States Department of Energy (DOE) indicate an even higher consumption up to 48% worldwide with peaks up to 70% of the electrical energy needs in Saudi Arabia [2-4]. Conventional cooling and heating technologies, such as air conditioning, are affected by a relatively low efficiency, high energy consumption and significant environmental hazards [5]. As a result, passive methods for cooling and heating are in great demand. Daytime radiative cooling has recently emerged as passive source of cooling power using outer space as a cold sink by dumping thermal radiation through the transparent atmospheric window (8-13 µm) [6,7], driving a broad effort worldwide to model and develop new materials with high solar reflectance and thermal emissivity [5,8–17]. Due to the excellent cooling performance of the material, multifunctional radiative cooling materials are also receiving more and more attention, such as colored radiative cooling materials with patterned surfaces [18,19] ,radiation modulated electricity generation [20] and thermal camouflage [21,22].

Many radiative cooling materials, however, are characterized by a complex optical structure (including multilayer dielectric films, photonic crystals, and polymer-metal mirror composites), resulting in high cost, uncertain stability to weathering agents, and limited upscaling potential. In addition, due to their fabrication processes, it is often unfeasible to adapt these coatings to arbitrary shapes and large surfaces as often found in the building sector [11,23]. Paint-like mixtures represent a promising approach to overcome these issues, with several relevant examples in the recent literature [5,24,25]. For instance, Mandal et al. used a two-phase solvent separation method to prepare a porous polymer coating exhibiting a cooling power of 96 W m$^{-2}$ during the day and a sub-ambient temperature drop of 6 °C [5]. Despite the clear advantages in terms of scalability and their high cooling performance, these synthetic polymer coatings still present some undesirable disadvantages in terms of adhesion, cost, poor mechanical properties on top of their significant environmental impact [5,26–32]. For these reasons, developing passively cooling materials based on abundant biomass materials, which are inexpensive, environmentally friendly and with strong mechanical properties is of great interest.

As a typically sustainable material, wood, which is composed of cellulose, lignin, and hemicellulose, has been widely used in our daily life due to the low cost, good processability, and environmental friendliness [33–37]. Lignin is strongly absorbing in the visible spectrum, while the cellulose nanofibers are largely transparent in the solar spectral range while being highly emissive in the infrared range, making delignified wood an ideal material for radiative cooling applications [38–40]. This was recently demonstrated by Li et al., who reported on a delignification and densification strategy for the fabrication of a mechanically strong and radiative cooling wood-based structural material [41]. Similarly, Chen et al. used a bottom-up method to assemble inorganic microspheres into a cellulose substrate obtaining a high-strength radiative cooling wood with flame-retardant and antibacterial properties [42]. Other cellulose derivatives are also available for the preparation of radiative cooling materials [43]. Despite recent progress on the development of wood-based cooling materials, these examples are still unsuitable for many building application scenarios as the energy savings in summer can be easily displaced by even larger heating energy needs during the winter. Therefore, developing a bifunctional material compatible with both radiative cooling and solar heating remains highly desirable and yet technology challenging [44–48]. After delignification of the wood, more hydroxyl groups are exposed, which means that solvents containing hydroxyl groups will easily enter the material by hydrogen bonding. Meanwhile, as more pores are exposed, space is provided for solvent storage. These mean that finding a solvent containing hydroxyl groups and having the same refractive index as heterocellulose will make the realization of rapid conversion of the optical properties of the material possible.

In this study, we designed an optical wood with switchable radiative cooling and transmittance of solar irradiation through complete delignification and phenylethanol impregnation. Highly reflective wood (OW-I state) with solar reflectance of 94.9% in the visible can be obtained through complete delignification. After the absorption of phenylethanol (a naturally available alcohol posing no health or environmental hazard, often used as an edible flavor [49]), the solar transmittance of optical wood (OW-II state) can reach 68.4% for the same 3 mm thick sample (30 mm × 30 mm). As a result, the highly reflecting OW-I can achieve a sub-ambient temperature drop of 4.5 °C, under direct sunlight at an ambient temperature of 32 °C (May 22, 2021 in Harbin, China). For passive heating, the OW-I is filled with phenylethanol to reach a highly transmittance state with a solar reflectance of 31.6%, leading to a net solar heating (5.6 °C, November in Harbin) above the temperature of natural wood (NW) under the same irradiance conditions. The equivalent cooling and heating powers associated to the two states are estimated at 81.4 W m$^{-2}$ and 229.5 W m$^{-2}$, respectively, holding great promise for all-season energy savings in energy-efficient buildings.

## 2. Experimental Section

*2.1. Materials*

*Populus tomentosa Carr.* was cut into wood blocks of 30 mm × 30 mm × 3 mm (L × W × H). The wood blocks were washed with distilled water and absolute ethanol for 20 min, then dried to absolute dryness at 45 °C. Sodium chlorite (NaClO$_2$), glacial acetic acid (CH$_3$COOH) and phenylethanol were purchased from Aladdin Bio-Chem Technology Co., Ltd. (Shanghai, China). The ethanol and anhydrous sodium acetate (NaAc) were purchased from Fuyu Chemical Co., Ltd. (Tianjin, China). All other reagents and solvents were analytically pure.

*2.2. Characterization*

The morphologies of the samples were observed using a scanning electron microscope (SEM, JSM-7500F). Phase analysis of all samples was performed by means of XRD, (X'Pert$^3$

Powder, Cu-Kα radiation λ = 1.5405 Å, 2θ = 10°–60°, scan speed = 4° min$^{-1}$). FTIR (frontier) analyses was used to measure the chemical bonding of the sample. The UV-vis-NIR reflectivity was measured using a PerkinElmer Lambda 950 instrument. The specific surface area was measured using an Autosorb-iQ $N_2$ adsorption apparatus. The temperatures were measured using an AS887 K-type four-channel thermocouple temperature meter (Xima Instruments Co., Ltd., Dongguan, China). The contact angle was measured using an Optical Contact Angle Meter (Dataphysics-OCA20). The thermal conductivity of NW and OW-I wood were measured using a Thermal conductivity meter (Hot Disk TPS2500S).

*2.3. Preparation of delignified wood*

A 1.5 wt% $NaClO_2$ with acetate buffer solution (pH 4.6) was used as the lignin removal solution. The dried wood blocks were completely immersed in the lignin removal solution for 10 h under an absolute pressure of 10 kPa. Immersion at 80 °C for 9 h, with solution changes every 3 h, until the wood turned pale. The treated wood was then soaked in water for 24 h for purification. After being frozen at −12 °C overnight, the samples were then freeze dried at −45 °C for 24 h. Subsequently, the OW was obtained. The high-reflectivity state of OW is defined as OW-I. The transmittance state of OW is defined as OW-II.

The optical wood (as the OW-I state) was pre-impregnated with ethanol for 3 min to increase the wettability and shorten the immersion time of the whole process, followed by immersion in phenylethanol until completely transparent (as the OW-II state). The OW-II with high transmittance was immersed in ethanol until opaque, and then air-dried to reduce it to radiative cooling wood (OW-I).

## 3. Results and discussion

*3.1. Morphological characterizations of NW and OW*

Fig. 1 shows a conceptual roof including an OW layer for an energy-efficient building. The OW can be switched between radiative cooling and daytime heating to keep the indoor thermal comfort level without external energy consumption. Compared with other radiative cooling materials reported in the literature, the OW offers high mechanical strength, excellent flexibility, and switchable optical performances between a highly reflective cooling state and a highly translucent one for solar heating, thus offering potential energy savings during all seasons. The OW is composed of microporous cellulose fibers resulting in a highly diffuse reflection without secondary light pollution caused by direct reflections. Meanwhile, the haze of OW-II exceeds 90%, which can effectively protect internal privacy. The OW can be reversibly switched between the reflective state (solar reflectance 94.9%, OW-I) and the transparent state (solar transmittance 68.4% from 250 nm to 2500 nm, OW-II) through the absorption and desorption of phenylethanol (Fig. 1a). During summer, OW-I exhibits a net radiative cooling action (Fig. 1b, left), while in winter, OW-II allows sunlight to enter and keep indoor spaces warm (Fig. 1b, right).

Natural wood tends to absorb visible-wavelength light due to the presence of lignin. We can adjust the chemical compositions and microstructure of the wood cell wall to meet the requirements of high and broadband reflectance over the solar spectrum through delignification. With the reduction of lignin content from 25.90% to 1.33% (Fig. S1), the OW-I (Fig. 2a) becomes white due to the removal of color-emitting groups and major heat-absorbing groups of lignin. Similar X-ray diffraction (XRD) peaks (Fig. S2) were found in OW-I and NW, indicating that the delignification treatment does not destroy the crystal structure of cellulose, thus largely preserving the mechanical strength of the sample [50].

The three major components of cellulose, hemicellulose, and lignin are aligned and assembled in wood cell walls. As shown in Fig. 2b and c, there is a large amount of dense lignin in the intercellular layer binding the cells tightly to each other. After delignification, the hierarchal alignment of cellulose nanofibers is well-exposed in the cell wall of OW-I. Large pores and nanopores are exposed in the intercellular layer, and the neighboring cell walls become more loosened (Fig. 2d and e). This causes a doubling of the pore volume, a decrease of the average pore size from 7.64 nm to 6.10 nm, and an increase of specific surface area from 2.83 $m^2\ g^{-1}$ to 5.41 $m^2\ g^{-1}$ (Fig. 2g, Fig. S3 and Table S1). The scattering is increased due to the loosening of the sample structure and the removal of lignin (containing endothermic and chromophoric groups), resulting in a visibly white material.

Lignin contains a large number of conjugated structures (more details can be found in the Supporting material) containing π electrons with high activity and a low excitation energy required for their electronic transition. As a result, NW presents higher molar absorption coefficients than OW-I, resulting in its typical brown color and low reflection under most wavelengths in the solar spectrum. When lignin is removed from NW, the OW-I becomes white due to the optical scattering of cellulose. At the same time, the strong absorption peaks of the NW and the OW-I between 770 and 1250 $cm^{-1}$ are attributed to C–H bending vibration, and C–O, C–O–C tensile vibration of cellulose [51], falling exactly inside the transparent window range of the atmosphere. The high emissivity in the infrared range indicates that OW-I is a good radiative cooling material.

*3.2. Infiltration and conversion performance of NW and OW*

To transform the OW from its high-reflectance state OW-I to the high-transmission state OW-II, we investigated the wettability of OW-I toward phenylethanol, a natural and widely available food flavor which has a good compatibility with cellulose and similar refractive index. In the wettability and contact angle tests of phenylethanol, a 5 cm long OW-I sample was completely wet after a 90 s contact with the solvent, which reached only 16.5% of a NW sample of the same length, for comparison (Fig. 2i–l). Similarly, in the contact angle test, a solvent droplet is completely absorbed (0° contact angle) into the NW after 24 s, compared to the 1.2 s of the OW-I case (Fig. 2j and k). Such large difference is due to the fact that, after the removal of lignin from the wood cell wall, more micropores and hydroxyl-rich encapsulated cellulose are exposed to providing many active sites for hydrogen bonding (Fig. 2f–h), which improves the wettability of phenylethanol and cellulose favoring its capillary infiltration and enabling a fast optical switching.

In order to increase solar gains and passive heating of an indoor environment, the OW must possess a high optical transmittance. The rapid conversion of OW from a high solar reflectance (OW-I) to a high solar transmission state (OW-II) can be achieved by removing the optical scattering interfaces between cellulose and air through phenylethanol impregnation, due to their similar refractive index of ~1.53 (Fig. 3a) [52,53]. The hydroxyl groups and microscopic structure of cellulose further facilitate the rapid and complete infiltration of phenylethanol endowing OW-II with its high optical transmittance. Fig. 3b–d shows the results of impregnating OW with phenylethanol. The transparency of OW increases significantly with the degree of impregnation (thick blue arrow). Notably, the transition can be perfectly reversed by rinsing the OW-II in ethanol (Fig. S5). Due to their hydroxyl groups (which are more likely to form hydrogen bonds), small size and high solubility to phenylethyl alcohol, ethanol molecules are extremely effective at washing the phenylethanol molecules out of OW, thus restoring the original OW-I state (thick red arrow). Under an optical microscope, OW-I shows many empty vessels in the initial state (thin red arrow in Fig. 3e), which are responsible for its strong scattering in the solar range. When in contact with phenylethanol, the catheter is filled (thin blue arrow), and the OW begins is optical switching (Fig. 3f). Finally, when the catheter is fully filled, the transition is complete and we obtain the high solar transmittance state (OW-II, Fig. 3g). The complete infiltration of phenylethanol in OW-II is confirmed by FTIR spectral analysis, as shown in the Supplementary material (Fig. S6). So, the empty microchannels of OW represent a stable storage space for the phenylethanol, effectively preventing its evaporation due to solar heating in winter. Throughout the transformation, the cellular structure is also not swollen, which allows OW-II to maintain the mechanical strength and flexibility of OW-I.

Rayleigh scattering formula was used to investigate the scattering of wood-switching process

$$I = \frac{24\pi^2 E_0^2 N V^2}{\lambda^4} \left(\frac{n_2^2 - n_1^2}{n_2^2 + 2n_1^2}\right) \qquad (1)$$

where $I$ is the total amount of light energy scattered per unit volume by the system under study, $E_0$ is the amplitude of the incident light, $\lambda$ as the wavelength of the incident light, $n_1$ and $n_2$ are the refractive index of dispersion medium and phase respectively, and the refractive index of air is 1, heterocellulose is 1.53, and phenylethanol is 1.5323, $N$ is the number of particles per unit volume, and $V$ is the volume of a single particle. When phenylethanol enters OW-I, the difference of refractive index becomes $n_{heterocellulose}^2 - n_{phenythanol}^2$, so its scattering intensity is almost zero and OW-I becomes OW-II.

Moreover, OW-I and OW-II both have higher tensile strengths than those reported in the literature for polymer-based radiative cooling materials, reaching 22.30 and 18.39 MPa, respectively (Fig. S7d) [5,54]. Around and axis parallel to the fiber orientation, and in contrast with the NW case, both OW-I and OW-II also show excellent flexibility thanks to their compressible micropores (Figs. S7a–c), reaching a minimum curvature radius of just 4 mm under bending. The combination of mechanical strength and flexibility makes this green, sustainable, low-cost material suitable for a wide range of applications.

*3.3. Optical and thermodynamic properties of NW and OW*

We measure the reflectance of OW-I, OW-II, and NW over a broad wavelength range using an ultraviolet/visible/near-infrared spectrometer. The average reflectance of NW is 20% in the visible range (380–780 nm), and solar reflectance is essentially 32.3% across the entire measured range (Fig. 3h). The solar reflectance of the OW-I is 94.9% in the visible, and above 90% in the range of 250–2500 nm, where the spectral energy is most concentrated. The solar reflectance of the OW-II was below 30% due to the penetration of phenylethanol, which results in a dramatic suppression of light scattering from the sample. Consequently, the OW-II shows broadband transparency corresponding to a solar transmittance of 68.4% between 250 nm and 2500 nm. The high solar transmittance allows the OW-II to raise the interior temperature by allowing more sunlight to enter the house, while the high haze (94%) guarantees for even heating and privacy (Fig. S8, Fig. S9, Video 1–2). The emission spectrum of OW-I is in the infrared range of 2.5–25 μm, which covers fully the wavelength range of room temperature black bodies (Fig. 3i), exhibiting a high emissivity (0.93) which allows it to radiate part of its thermal energy to outer space the atmospheric transparency window.

We perform a first comparison between the temperatures of different samples using a thermal camera (Fig. 4a and b and Fig. S10). Fig. 4a and b is thermal image of NW, OW-I, OW-II after 0 and 60 min of direct illumination, respectively. After 60 min, the temperatures of NW, OW-I, OW-II are 35.5 °C, 25.6 °C, 33.0 °C, respectively. The temperature of OW-I is about 7.5 °C lower than the marble floor temperature, suggesting a significant radiative cooling effect. The temperatures reached by NW and OW-II are higher than that of OW-I by 9.9 and 7.4 °C, respectively. This result proves that both wood and OW have the ability to heat using sunlight. However, the heating of wood is achieved by absorbing sunlight, while the OW-II exploits secondary absorption of transmitted sunlight. We verified the data of OW-I, NW, and OW-II by measuring the temperatures of the samples, and the test setup is shown in Figs. S11 and S12. The temperature of OW-I (Fig. 4c, Fig. S13) was lower than the temperatures of ambient air and NW (at ambient temperature 32 °C), with an average temperature difference with the ambient air of 4.5 °C (daytime) and 5.3 °C (nighttime). This demonstrates the radiative cooling effect of OW-I. In Table S3, Table S4 and Fig. S14, we demonstrate the reproducibility of the experimental results and quantified statistically significant differences between experimental conditions by *t*-test.

*3.4. Thermodynamic calculations of NW and OW*

In order to investigate the temperature test data, the cooling power of the material was calculated. In Fig. 4e and Fig. S16, the calculated radiative cooling power of OW-I is expected to deliver 81.4 W m$^{-2}$ at 25 °C. In that case, the calculated temperature of the OW-I is 7.5 °C lower than that of the ambient temperature (25 °C) at $h_{cc}$ = 6 W m$^{-2}$ K$^{-1}$ (ambient wind speed = 6 m s$^{-1}$). The theoretical calculation of the cooling temperature of the OW-I at 7.5 °C is consistent with the actual measured value of 7.5 °C (Fig. S13). The theoretically calculated NW temperature difference is 24.8 °C instead of the measured value of 17 °C, which is probably a consequence of the thermal insulation properties of wood. The thermal conductivity of NW and OW-I wood were measured using a Thermal conductivity meter, and fed into heat transfer calculations for the NW and OW-I samples, returning temperature differences of 24.7 °C and 10.5 °C with respect to ambient temperature, respectively (Notes S3), in good agreement with the theoretically calculated values. Fig. 4f shows heating power of OW-II up to 229.5 W m$^{-2}$ on the winter solstice in Harbin.

To simulate the performance of the OW as a roofing material, NW, OW-I, and OW-II elements were modeled into tiny houses with polypropylene figures placed inside (Fig. 5a and b). In winter (Fig. 4d, Fig. S15), we conducted a 60 min internal temperature monitoring of the house models at an ambient temperature of 3 °C (28 Nov 2021 in Harbin, China). Placed outdoors, the model cools down quickly, reaching a stable temperature after 15 min. The three models showed obvious temperature differences, with an asymptotic temperature inside the small model of around 1–2 °C for the OW-I roof, which is at most about 5.5 °C lower than NW and therefore unsuitable for the cold season. On the other hand, the indoor temperature of the OW-II roof is about 10 °C, up to 5.6 °C higher than that of the NW model, indicating the significance of the OW-II for winter use.

In summer, the systems were exposed to direct sunlight for 120 min (Fig. 5a–d). We found that the temperature of the figure inside the OW-I house reached 39.5 °C, which is 2.8 °C lower than that of the figure in the NW. On the other hand, the temperature of the figure in the OW-II house reached 48.5 °C, and the temperature gradient may make it as a roofing material. By Figs. S19–S21, the solar reflectance and IR emissivity of the material did not decrease significantly after 10 cycles, and the maximum solar reflectance did not change more than 2%. Both states still have good optical properties after long placement, and the maximum reflectance of OW-I only decreases by 4% after two years of placement. This indicates that the material has excellent cycling performance and long lifetime. The life cycle assessment shows that the material has a low environmental impact in all indicators during the cycle (Tables S5 and S6) [55,56]. An illustrative building roof module (500 mm × 500 mm) was also fabricated to show the potential upscaling of the material. The house roof module is shown in both states in Fig. 5e and f, with a pattern below the material being clearly visible in the transparent state. In the supporting material, we present additional comparisons of OW against other materials presented in the literature, showing its competitive properties in terms of overall cost (Fig. S22, Notes S4 and Table S7), but also tensile strength, cooling power, environmental sustainability, heating power and switchability.

## 4. Conclusions

In summary, we presented a smart thermal management strategy using a porous delignified wood material, which can switch between a highly solar reflective and solar transmission state via phenylethanol absorption and desorption to achieve the effect of heating in winter and cooling in summer. The fabrication of OW is simple, sustainable, it does not require environmentally harmful solvents, and has a very low fabrication cost compared to other passively cooling materials. At the same time, the OW retains good mechanical properties and flexibility in the high reflection state. In the transmission state, the covered space is uniformly illuminated by sunlight and yet invisible from the outside. For these reasons, we envision that OW represents a promising and versatile material for multiple applications related to energy efficiency of buildings.

**CRediT author contribution statement**

**He Gao:** Writing – original draft, Investigation, Data curation, Conceptualization. **Ying Li:** Writing – review & editing. **Yanjun Xie:** Writing – review & editing. **Daxin Liang:** Supervision, Investigation, Funding acquisition. **Jian Li:** Writing – review & editing, Visualization. **Yonggui Wang:** Writing – review & editing. **Zefang Xiao:** Writing – review & editing. **Haigang Wang:** Writing – review & editing. **Wentao Gan:** Writing – review & editing, Visualization. **Lorenzo Pattelli:** Supervision, Investigation, Funding acquisition. **Hongbo Xu:** Supervision, Investigation, Funding acquisition.

**Declaration of competing interest**

The authors declare that they have no known competing financial interests or personal relationships that could have appeared to influence the work reported in this paper.

**Data availability**

No data was used for the research described in the article.

**Acknowledgements**

Hongbo Xu acknowledges support from the Key Research and Development Program of the Ministry of Science and Technology under Grants No 2023YFB4604100. Daxin Liang and Hongbo Xu acknowledge support from the National Natural Science Foundation of China (No 31890772, 51702068, 52072096). Lorenzo Pattelli acknowledges support from the European project 21GRD03 PaRaMetriC. The project 21GRD03 PaRaMetriC received funding from the European Partnership on Metrology, co-financed by the European Union's Horizon Europe Research and Innovation Programme, and from the Participating States.

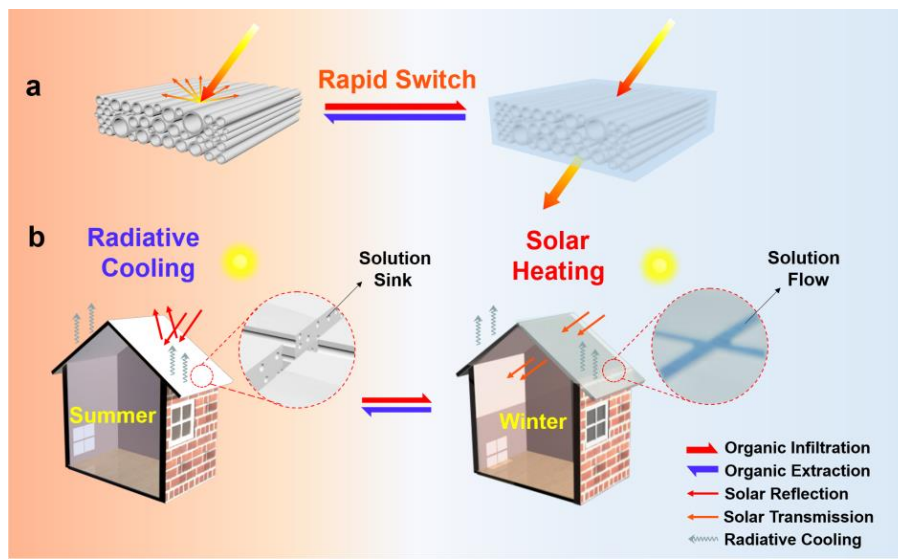

**Fig. 1.** Schematic diagram. (a) OW's rapid switching: OW-I (left) and OW-II (right), (b) The working principle of the energy-efficient building made with an OW roof, based on the controlled impregnation and rinsing of the wood panels with either phenylethanol (OW-I to OW-II) or ethanol (OW-II to OW-I).

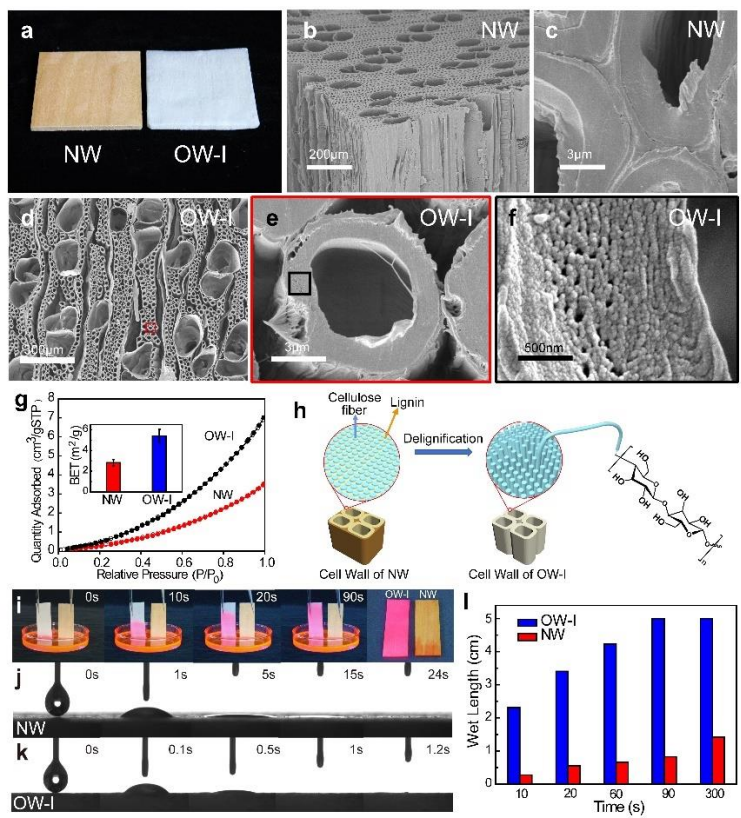

**Fig. 2.** Morphological characterization and wettability of as-prepared wood. (a) Optical images of NW (left) and OW-I (right), (b) Three-sectional SEM image of NW, (c) Cross-sectional SEM image of NW, (d–f) Cross-sectional SEM image of OW-I, (g) BET of NW and OW-I, (h) Schematic diagram of cell wall and the OW-I with nanofibril bundles and cellulose molecules, (i) Samples were tested for natural wettability to phenylethanol (stained with phenylethanol for visualization), Hydrophobic angle (solvent phenylethanol): (j) NW; (k) OW-I, (l) Infiltration rate as a function of infiltration time, where the data were obtained by natural wettability testing of NW and OW-I on phenylethanol.

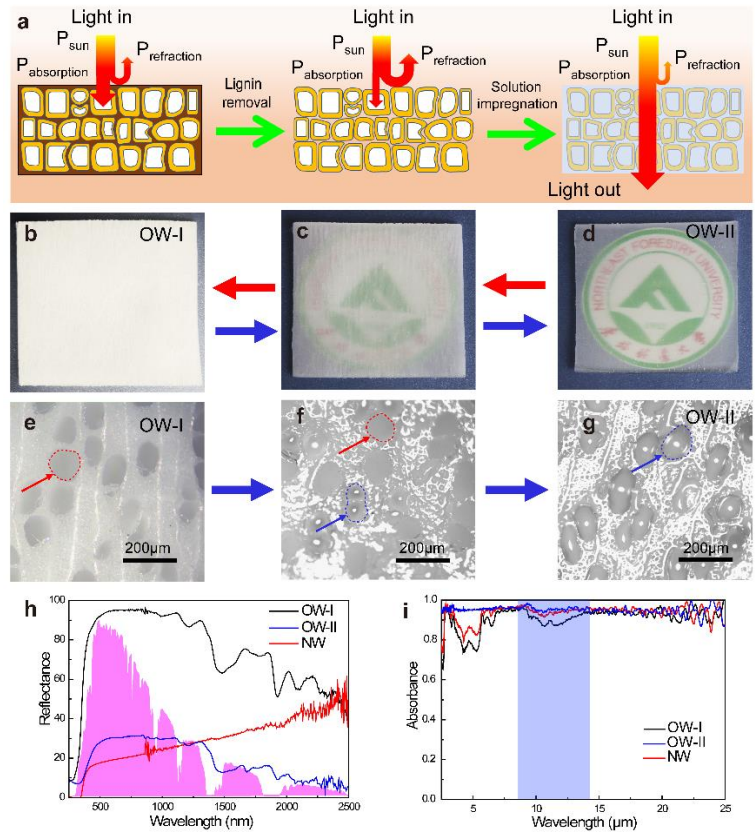

**Fig. 3.** Optical properties of wood. (a) Schematic diagram of light radiation on NW (left), OW-I (middle), and OW-II (right). Optical image of the conversion process, the blue arrow indicates the phenylethanol impregnation process and the red arrow indicates the ethanol rinsing process: (b) OW-I, (c) Partially impregnated wood, and (d) OW-II; Optical microscope image of the conversion process, the blue arrow indicates the phenylethanol impregnation process: (e) OW-I, (f) Partially impregnated wood, and (g) OW-II; (h) The reflection spectra of NW, OW-I and OW-II. The AM 1.5 solar spectrum (shaded in pink) is superimposed for reference. (i) Infrared emissivity spectra of NW, OW-I and OW-II between 2.5 and 25 μm. The blue shaded area highlights the atmospheric transparency range.

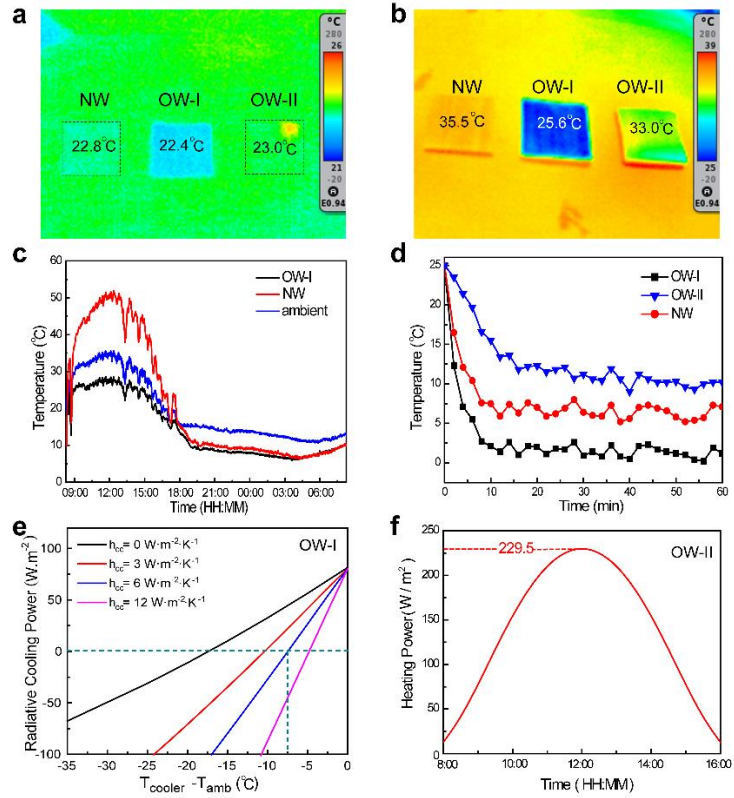

**Fig. 4.** Thermal management performances of OW. (a) Thermal infrared images of NW (left), OW-I (middle), and OW-II (right) on a marble slab indoors. (b) Thermal infrared images of NW (left), OW-I (middle), and OW-II (right) on a marble slab after 120 min outdoor exposure to direct sunlight. (c) A continuous 24-h (May) outdoor cooling test of NW (red line) and OW-I (black line). (d) The 60 min temperature of NW (red line), OW-I (black line), and OW-II (blue line) measured inside the model rooms. (e) Calculated cooling temperature as a function of radiative cooling power for OW-I. (f) Calculated heating power curve of OW-II in transparent state under direct sunlight on the winter solstice in Harbin by daylight harvesting.

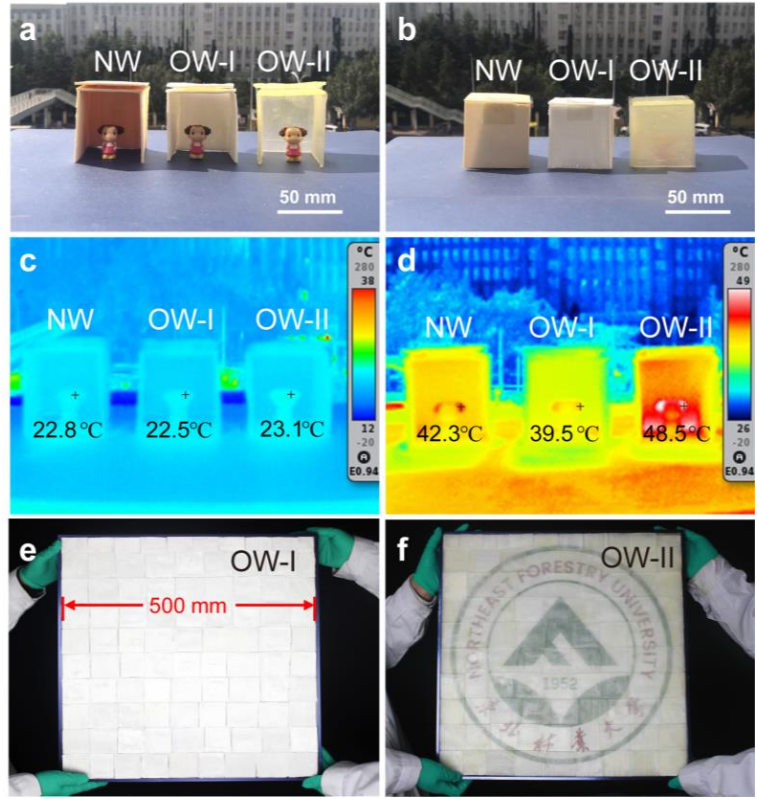

**Fig. 5.** House models built of OW. Optical images of house models built of NW (left), OW-I (middle), and OW-II (right). (a) Open state and (b) Closed state. Thermal infrared images of the house model which were briefly opened to allow the thermal imaging measurement: (c) at the beginning and (d) after irradiation for 120 min, the door is closed during irradiation. Photographs of the wood modules (assembled from 100 pieces of 50 mm × 50 mm samples) of (e) OW-I and (f) OW-II.

# Supporting Material for

# Optical wood with switchable solar transmittance for all-round thermal management

*He Gao[#ab], Ying Li[#c], Yanjun Xie[a], Daxin Liang[\*a], Jian Li[b], Yonggui Wang[a], Zefang Xiao[b], Haigang Wang[b], Wentao Gan[\*b], Lorenzo Pattelli[\*d,e], and Hongbo Xu[\*f]*

e-mail: daxin.liang@nefu.edu.cn, iamxhb@hit.edu.cn, wtgan@nefu.edu.cn, pattelli@lens.unifi.it

[a] Key Laboratory of Bio-based Material Science & Technology (Northeast Forestry University), Ministry of Education, Hexing Road 26, Harbin, 150040, P.R. China

[b] Engineering Research Center of Advanced Wooden Materials (Northeast Forestry University), Ministry of Education, Hexing Road 26, Harbin, 150040, P.R. China

[c] State Key Laboratory of Extreme Photonics and Instrumentation, ZJU-Hangzhou Global Scientific and Technological Innovation Center, Zhejiang University, Hangzhou, 310027, 310027, P.R. China

[d] Istituto Nazionale di Ricerca Metrologica (INRiM), Turin, 10135, Italy

[e] European Laboratory for Non-linear Spectroscopy (LENS), Sesto Fiorentino, 50019, Italy

[f] MIIT Key Laboratory of Critical Materials Technology for New Energy Conversion and Storage, School of Chemistry and Chemical Engineering, Harbin Institute of Technology, Harbin, 150001, P.R. China.

# These authors contributed equally to this work.

## Contents

**Figure S7**. (a-c) The optical image of OW-I flexibility. (d) Tensile properties of OW-I (red curve) and OW-II (black curve).

**Figure S8.** The UV-vis transmittance and haze diagram of OW-II.
**Figure S9.** The image of light on OW-I and OW-II.
**Figure S10.** The optical image of NW (left); OW-I (middle); OW-II (right) on the Marble slab.
**Figure S11.** Schematic of measurement setup.
**Figure S12.** (a) Schematic of the test setup. Temperature of OW-I (left) and NW (right). The input/output energy balance is marked with $P_{rad}$, $P_{sun}$, $P_{atm}$, and $P_{con}$, denoting the radiated power from the cooler, absorbed power from the sun, absorbed power from the atmosphere, and conduction/convection power loss, respectively. (b) Optical image of the test setup.
**Figure S13.** A cooling test continuous 24 hours (May): the black point indicates the temperature difference between OW-I and air, and the red point indicates the temperature difference between NW and air.
**Table S3.** Cooling test at OW-I's daytime highest temperature on different days in Harbin.
**Table S4.** Significant difference analysis between OW-I and ambient temperature.
**Figure S14.** Histogram of significant difference analysis between OW-I and ambient temperature.
**Figure S15.** The 60-minute indoor temperature (Nov. 28) of NW (red line), OW-I (black line) and OW-II (blue line) model room: temperature (top) and temperature difference (bottom).
**Figure S16.** The calculated temperature. (a) Heating temperature of NW. (b) Cooling temperature of OW-I.
**Figure S17.** The solar irradiance over the 9 h experimental period in May.
**Figure S18.** The calculation of 24 h solar irradiance on winter solstice in Harbin.
**Figure S19.** (a) The UV-vis reflection spectra and (b) Infrared emissivity spectra 10 optical performance switching cycles.
**Figure S20.** Optical photo of OW-II after 1 month of outdoor exposure.
**Figure S21.** (a) The UV-vis reflection spectra and (b) Infrared emissivity spectra.
**Table S5.** EIs results for the switching cycle of optical properties of OW.
**Table S6.** Percentage of phenylethyl alcohol and ethanol in EIs for the switching cycle of optical properties of OW.
**Notes S1.** Numerical models for radiative cooling.
**Notes S2.** Numerical models for solar heating.
**Notes S3.** Calculation of temperature difference due to sample thickness.
**Notes S4.** Comparison of the fabricated cost of optical wood and the current methods reported in the literature.
**Figure S22.** The radar map composed of optical wood and current methods reported in the literature.
**Table S7** Comparison of six character of optical wood and current methods reported in the literature.

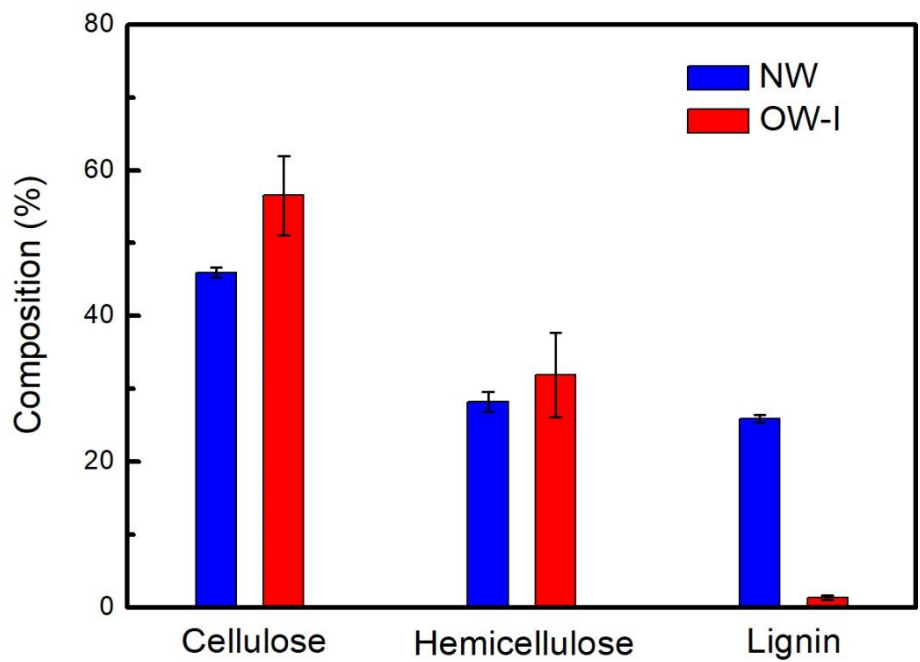

**Figure S1.** The composition of cellulose, hemicellulose and lignin in NW and OW-I.

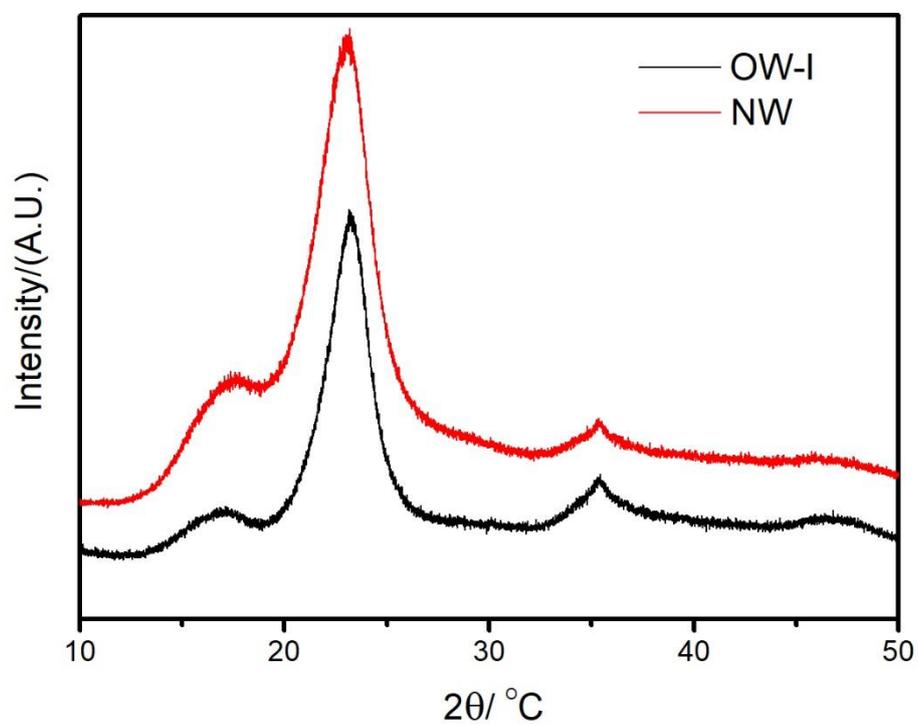

**Figure S2.** The XRD diagrams of NW and OW-I.

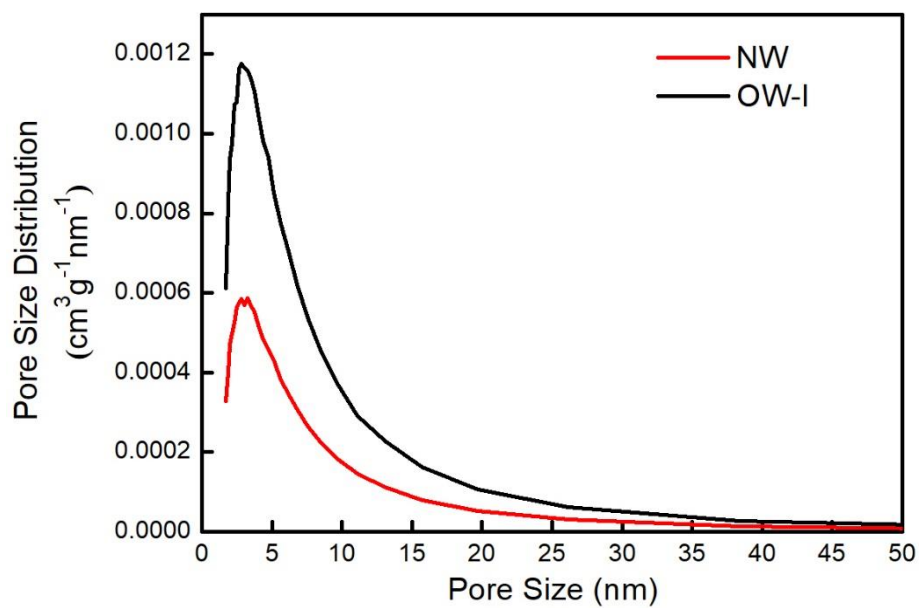

**Figure S3**. Pore size distribution curve of NW and OW-I.

Table S1 Channel structure properties of adsorbents

| Adsorbent | Surface area/ m² g⁻¹ | Pore Volume/ cm³ g⁻¹ | Pore Size/ nm |
|---|---|---|---|
| NW | 2.83 | $5.32 \times 10^{-3}$ | 7.651 |
| OW-I | 5.41 | $1.74 \times 10^{-2}$ | 6.10 |

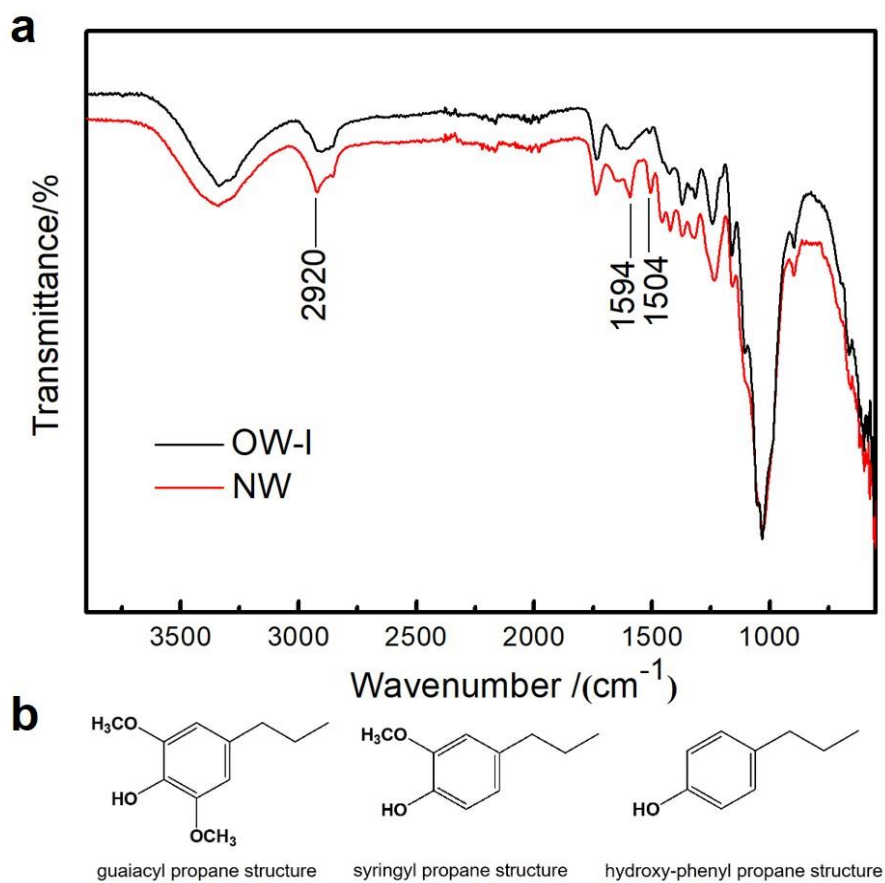

**Figure S4.** (a) The FT-IR of NW and OW-I. (b) Three monomer structures of lignin.

**Table S2.** The maximum absorption wavelength $\lambda_{max}$ and the corresponding molar absorption coefficient $\varepsilon$ of the chromophore group.

| Chromophore group | | $\lambda_{max}$/nm | $\varepsilon$/(L mol$^{-1}$ cm$^{-1}$) | $\lambda_{max}$/nm | $\varepsilon$/(L mol$^{-1}$ cm$^{-1}$) |
|---|---|---|---|---|---|
| Ether group | -O- | 185 | 1000 | | |
| Olefin | C=C | 190 | 8000 | 270-285 | 18-30 |
| Ketone | C=O | 195 | 1000 | 202 | 6900 |
| Benzene | | 184 | 46700 | | |

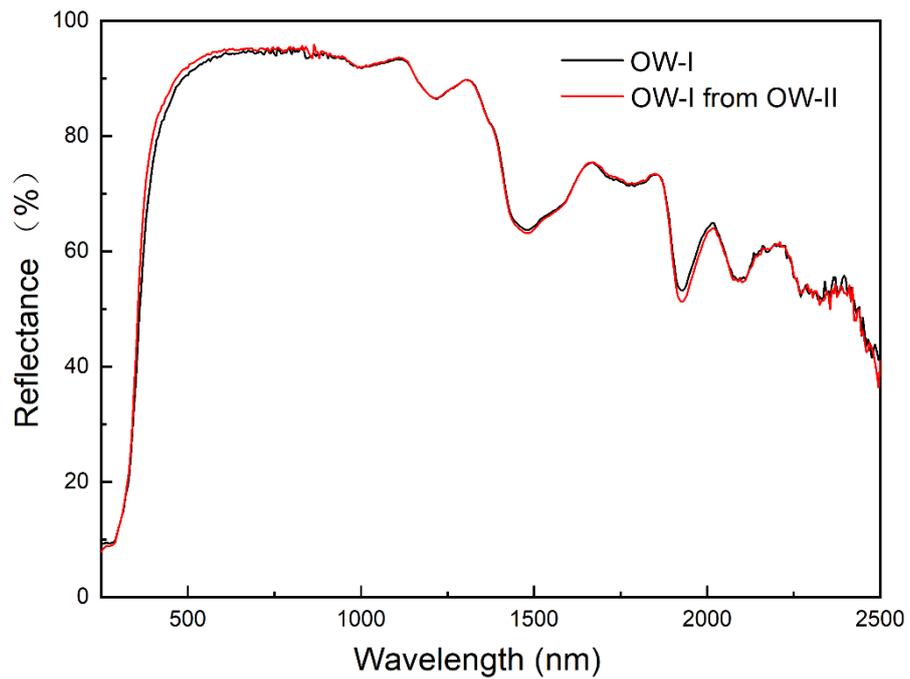

**Figure S5.** The UV-vis reflection spectra of OW-I and OW-I from OW-II after ethanol.

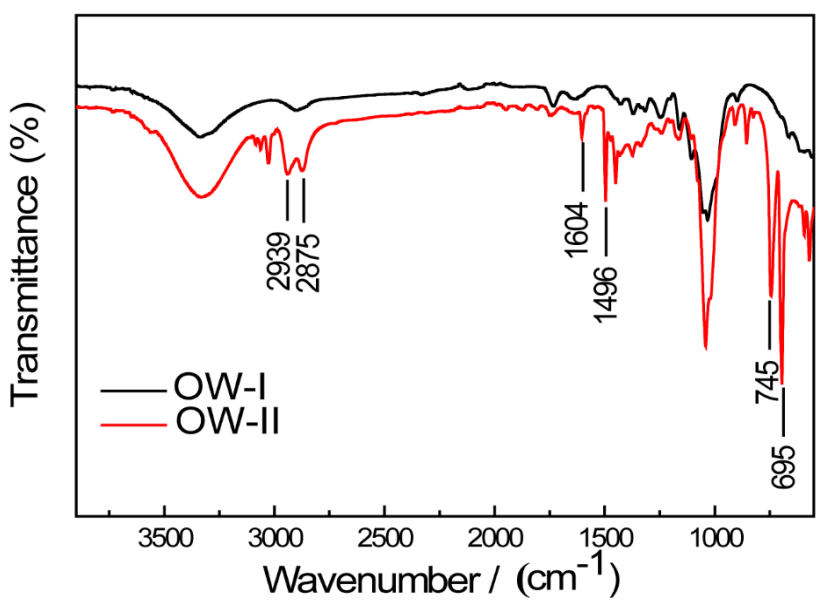

**Figure S6.** The FT-IR of OW-I and OW-II.

The main structural unit of lignin is phenylpropane, and there are three basic structures: guaiacyl, syringyl, and p-hydroxyphenyl structures (**Figure S4b**). We used Fourier-transform infrared spectroscopy (FTIR) to confirm the degradation of lignin. For the NW (**Figure S4a** red line), the extra absorption peaks at 2920 and 2853 cm$^{-1}$ are attributed to-CH$_2$-symmetric stretching vibration and asymmetric stretching vibration, respectively, and the absorption peaks at 1594 and 1504 cm$^{-1}$ belong to C=C stretching vibrations on the aromatic ring skeleton due to the lignin structure. After delignification, the characteristic peaks of lignin monomers (e.g., methylene groups, methyl groups, and phenolic hydroxyl group) in OW-I disappear (**Figure S4a** black line), indicating the removal of lignin from the wood. The lignin macromolecule contains more chromogenic groups, such as the benzene ring ( 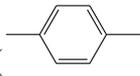 ), carbonyl group ( 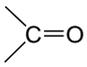 ), vinyl group (-CH=CH$_2$), and coniferylaldehyde group ( 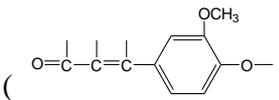 ), when these monomers are combined. These groups (**Table S2**) of the lignin are all chromogenic groups with large molar absorption coefficients. The coniferylaldehyde group is a large chromogenic group containing a C=O and conjugated C=C structure.

In **Figure S6**, the extra absorption peaks of OW-II at 2939 and 2875 cm$^{-1}$ are attributable to the -CH$_2$- symmetric and asymmetric stretching vibrations and the extra absorption peaks at 1604 and 1496 cm$^{-1}$ are attributable to the C=C stretching vibrations on the aromatic ring skeleton. The other absorption peaks of OW-II at 745 and 695 cm$^{-1}$ are attributable to the out-of-plane bending vibrations of Ar–H caused by the monosubstituted benzene ring, and the additional absorption peaks from the impregnated phenylethanol, and this confirms the complete impregnation of OW-I with phenylethanol.

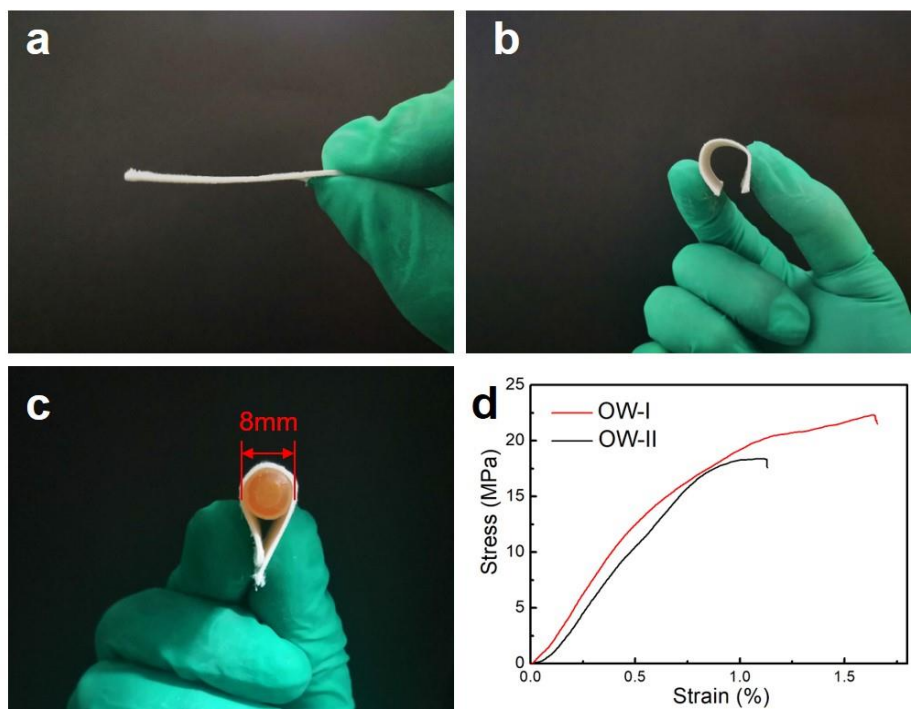

**Figure S7**. (a-c) The optical image of OW-I flexibility. (d) Tensile properties of OW-I (red curve) and OW-II (black curve).

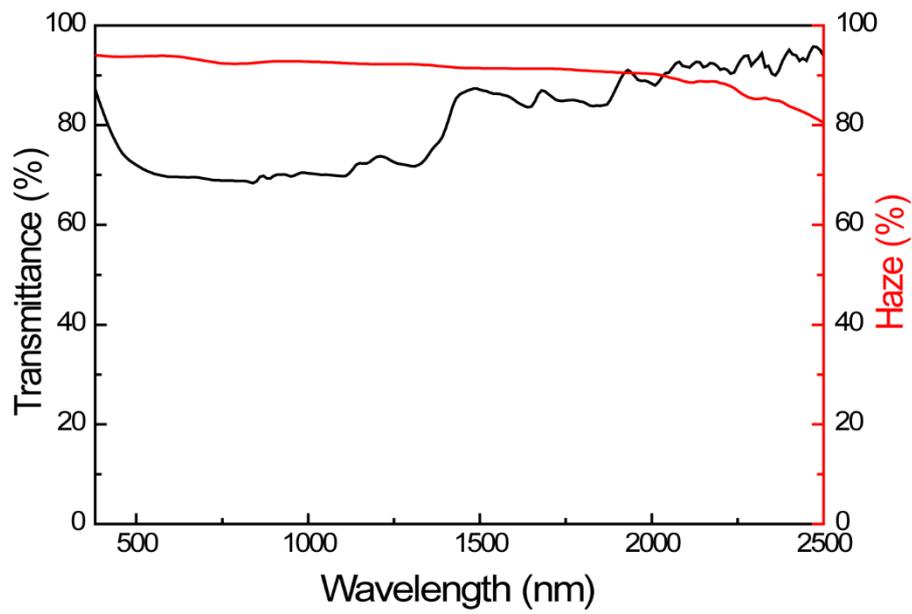

**Figure S8.** The UV-vis transmittance and haze diagram of OW-II.

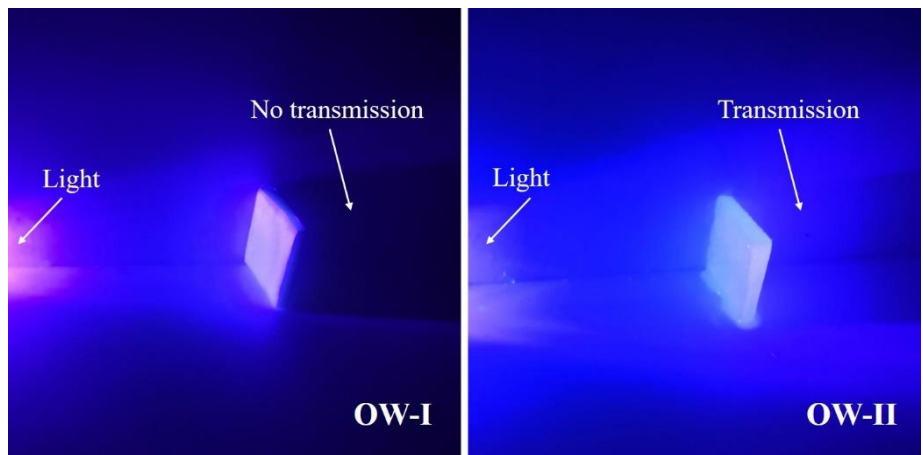

**Figure S9.** The image of light on OW-I and OW-II

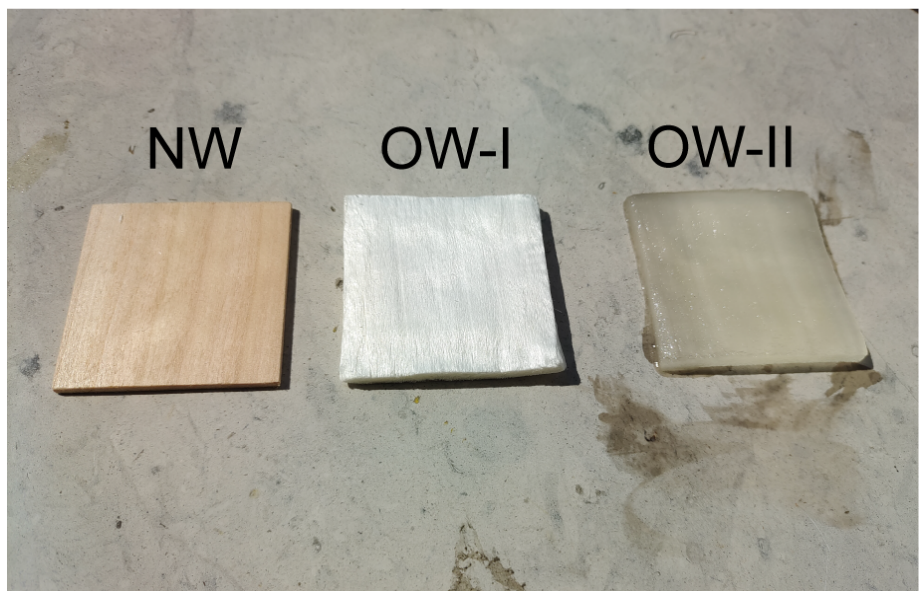

**Figure S10.** The optical image of NW (left); OW-I (middle); OW-II (right) on the Marble slab.

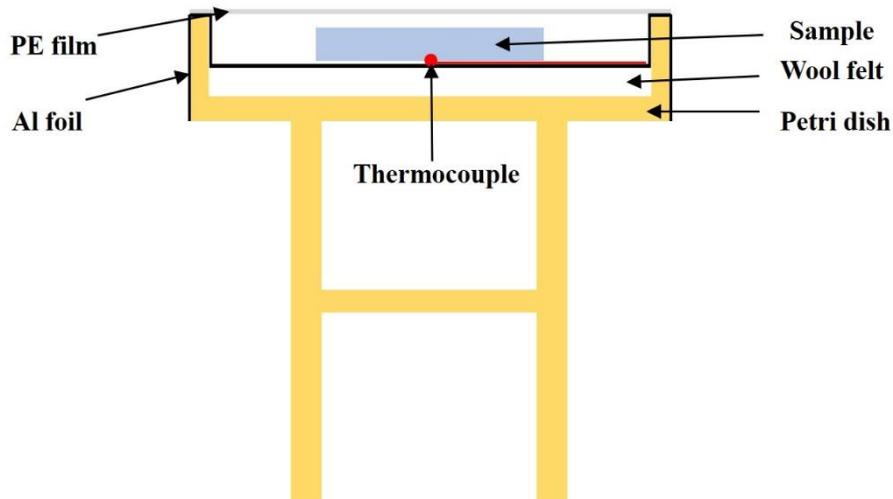

**Figure S11**. Schematic of measurement setup.

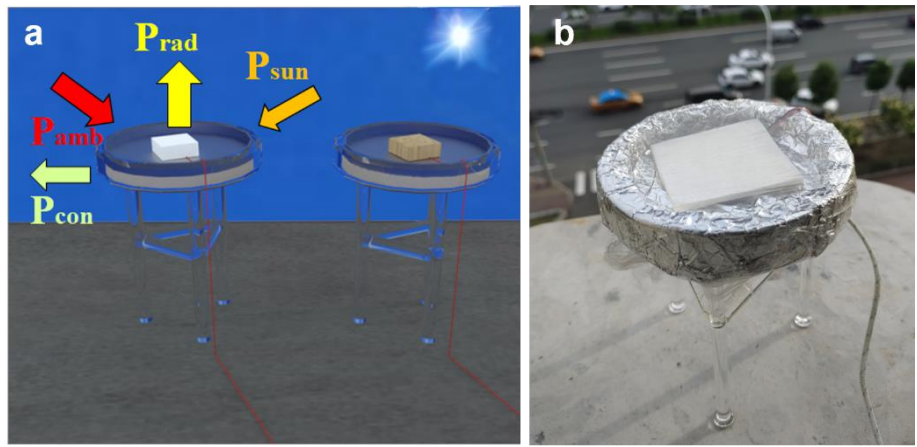

**Figure S12.** (a) Schematic of the test setup. Temperature of OW-I (left) and NW (right). The input/output energy balance is marked with $P_{rad}$, $P_{sun}$, $P_{atm}$, and $P_{con}$, denoting the radiated power from the cooler, absorbed power from the sun, absorbed power from the atmosphere, and conduction/convection power loss, respectively. (b) Optical image of the test setup.

The performance of the optical wood was tested on the sky-facing building roof of a building in Harbin, (45°43′49″N, 126°38′11″E), by exposing it to the sky. A schematic of the setup is shown in **Figure S11** and **Figure S12**. To achieve cooling below ambient temperature experimentally, special care needs to be taken in the measurement setup to reduce the heat conduction and convection from the ambient air. In our measurement, a Petri dish containing a wool felt is supported by three glass rods to suspend it above the roof and reduce non-radiative heat losses from ambient air. The

Petri dish is further covered with an aluminum foil to screen its infrared emission, with the sample placed on top of the foil and covered with a polyethylene film is placed over the Petri dish and the wool felt to reduce glass infrared radiation. The sample is placed on the aluminum foil, then the top of the Petri dish is covered by a polyethylene film, acting as a convection shield that is transparent to all the radiative wavelengths of interest. The temperatures of the device and ambient air are recorded using K-type thermocouples.

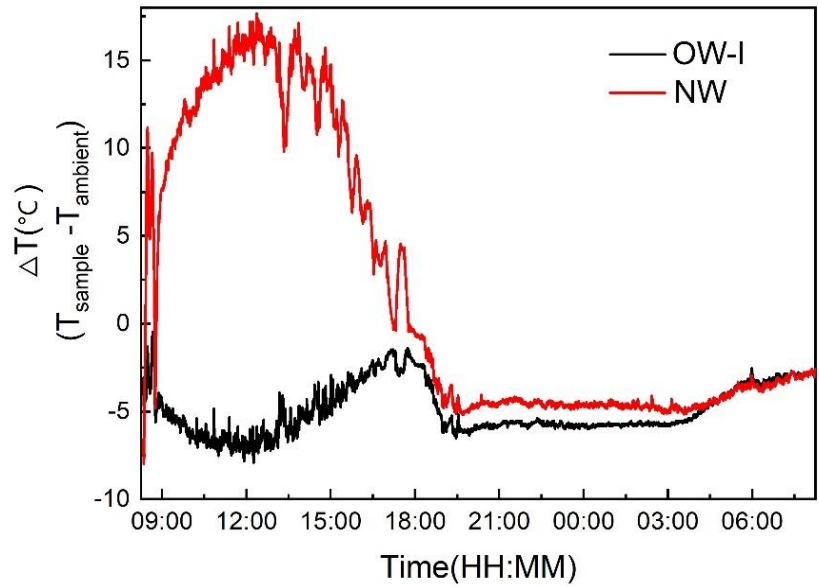

**Figure S13.** A cooling test continuous 24 hours (May): the black point indicates the temperature difference between OW-I and air, and the red point indicates the temperature difference between NW and air.

**Table S3.** Cooling test at OW-I's daytime highest temperature on different days in Harbin.

| dates | May 22, 2021 | May 17, 2021 | Sept. 10, 2020 | Aug. 21, 2020 | June 29, 2020 |
|---|---|---|---|---|---|
| ambient temperature (°C) | 32.0 | 28.5 | 28.9 | 29.3 | 28.8 |
| OW-I temperature (°C) | 27.5 | 24.9 | 25.8 | 25.9 | 24.7 |

**Table S4.** Significant difference analysis between OW-I and ambient temperature.

| Tukey's comparisons test | Mean Diff. | 95% CI of diff | Significant? | Summary | Adjusted Value |
|---|---|---|---|---|---|
| ambient vs. OW-I | -3.740 | -5.603 to -1.877 | Yes | ** | 0.0017 |

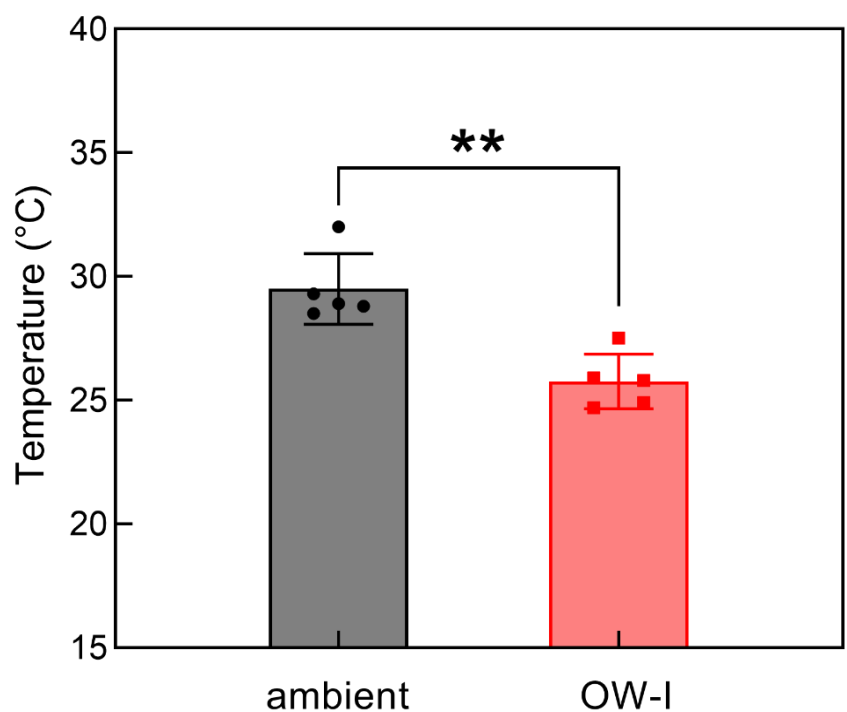

**Figure S14.** Histogram of significant difference analysis between OW-I and ambient temperature.

The asterisks (*) indicate statistically significant differences between the OW-I temperature compared to the ambient temperature determined by t-test (*$P < 0.05$; **$P < 0.01$).

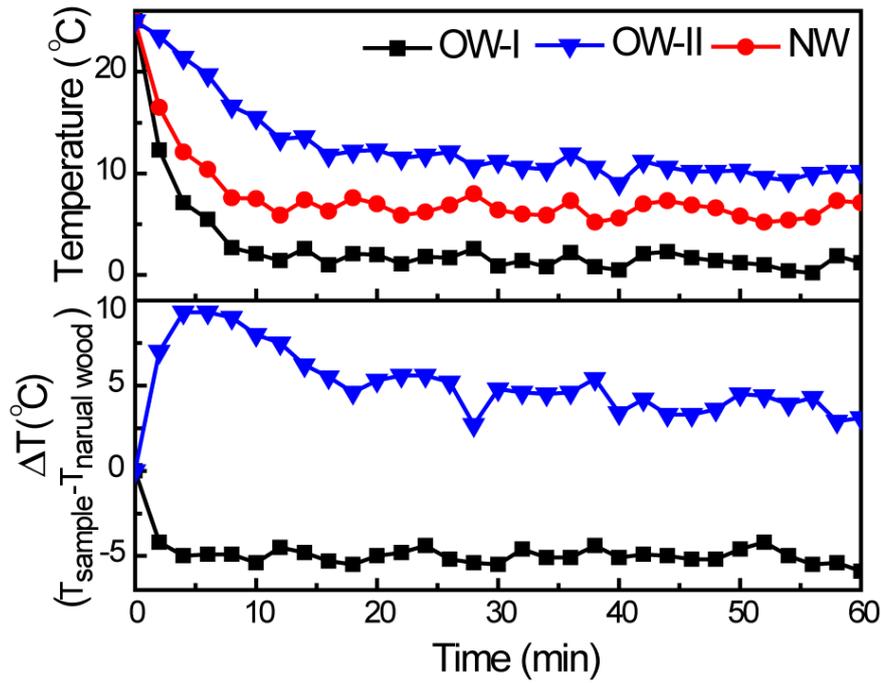

**Figure S15.** The 60-minute indoor temperature (Nov. 28) of NW (red line), OW-I (black line) and OW-II (blue line) model room: temperature (top) and temperature difference (bottom).

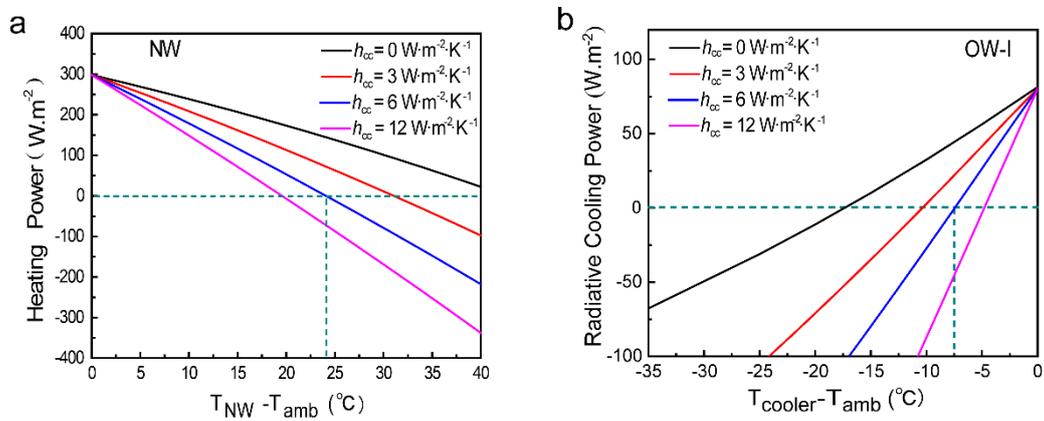

**Figure S16.** The calculated temperature. (a) Heating temperature of NW. (b) Cooling temperature of OW-I.

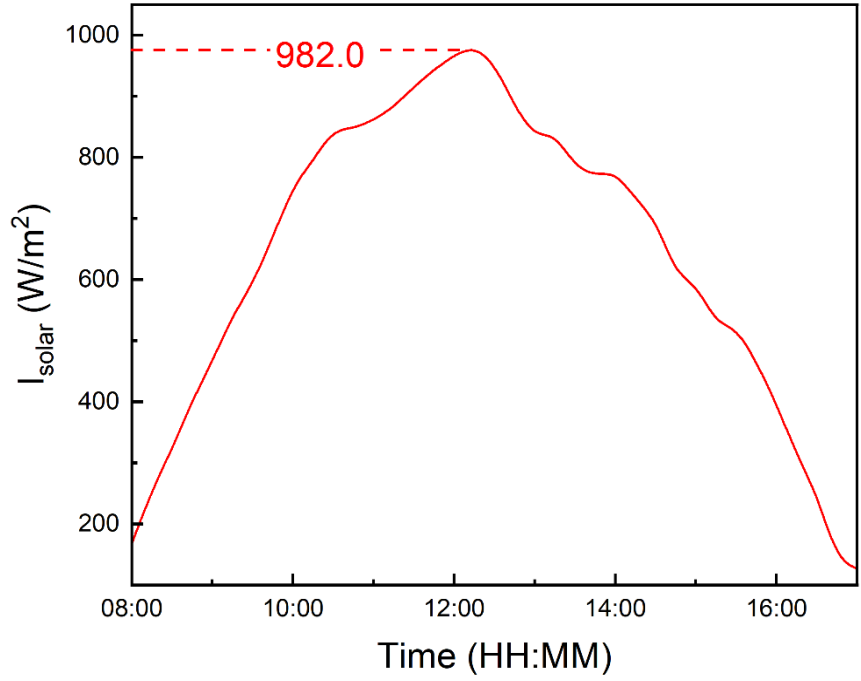

**Figure S17.** The solar irradiance over the 9 h experimental period in May.

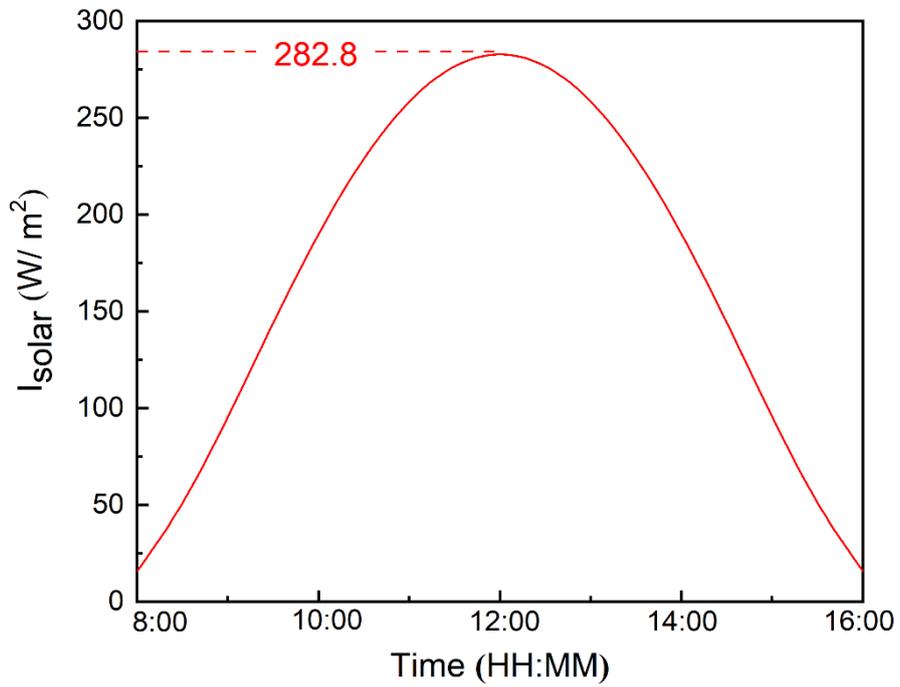

**Figure S18.** The calculation of 24 h solar irradiance on winter solstice in Harbin.

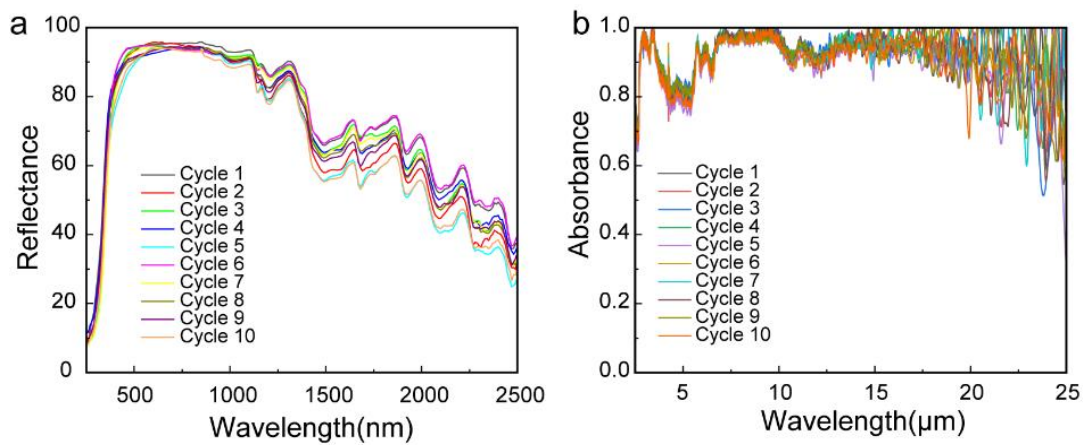

**Figure S19.** (a) The UV-vis reflection spectra and (b) Infrared emissivity spectra 10 optical performance switching cycles of OW.

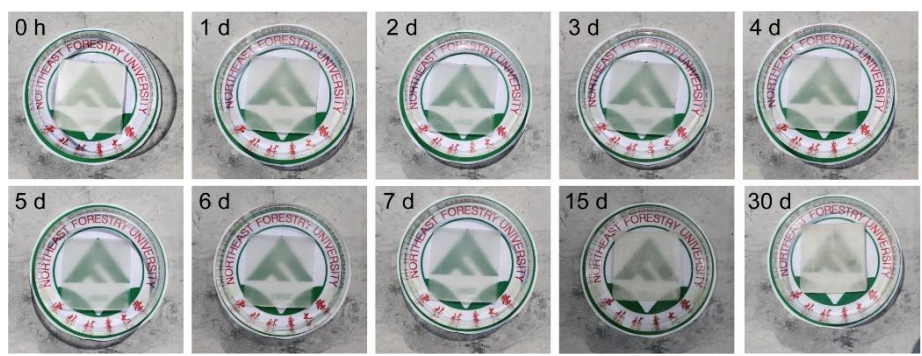

**Figure S20.** Optical photo of OW-II after 1 month of outdoor exposure.

We simulated the use of OW-II in natural outdoor weather and replaced the frame with a Petri dish during use. The material showed significant transparency even after being placed under direct sunlight, wind, etc. for one month.

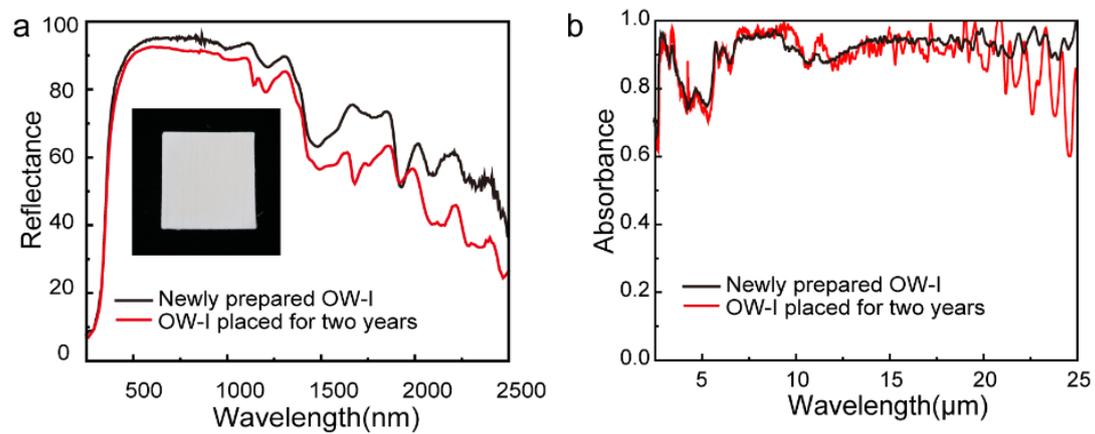

**Figure S21.** (a) The UV-vis reflection spectra and (b) Infrared emissivity spectra of OW-I.

**Table S5.** EIs results for the switching cycle of optical properties of OW.

| Categories | Unit | Data |
|---|---|---|
| AP | kg $SO_2$-Eq | 6.81E-05 |
| GWP | kg $CO_2$-Eq | 0.0201 |
| EP | kg $NO_x$-Eq | 4.04E-05 |
| FAETP | kg 1,4-DCB-Eq | 0.00244 |
| FSETP | kg 1,4-DCB-Eq | 0.00469 |
| HTP | kg 1,4-DCB-Eq | 0.00468 |
| IR | DALYs | 1.56E-11 |
| LU | $m^2a$ | 1.68E-05 |
| MA | $m^3$ air | 35.83237 |
| MAETP | kg 1,4-DCB-Eq | 0.0093 |
| MSETP | kg 1,4-DCB-Eq | 0.00882 |
| PCO | kg formed ozone | 2.26E-05 |
| R | kg antimony-Eq | 3.36E-04 |
| ODP | kg CFC-11-Eq | 3.03E-10 |
| TAETP | kg 1,4-DCB-Eq | 1.08E-06 |

**Table S6.** Percentage of phenylethyl alcohol and ethanol in EIs for the switching cycle of optical properties of OW.

| Categories | Phenethyl alcohol percentage (%) | Ethanol percentage (%) |
|---|---|---|
| AP | 68.33 | 31.67 |
| GWP | 66.43 | 33.58 |
| EP | 66.29 | 33.71 |
| FAETP | 60.85 | 39.35 |
| FSETP | 59.50 | 40.50 |
| HTP | 60.39 | 39.69 |
| IR | 84.55 | 15.45 |
| LU | 42.57 | 57.43 |
| MA | 27.41 | 72.59 |
| MAETP | 61.30 | 38.66 |
| MSETP | 60.02 | 40.02 |
| PCO | 14.26 | 85.74 |
| R | 64.53 | 35.47 |
| ODP | 22.05 | 77.95 |
| TAETP | 52.41 | 47.59 |

The goal of the life cycle assessment (LCA) is to analyze the environmental effects (EIs) of the switching cycle of optical properties of OW, where the environmental impact is the amount of phenylethyl alcohol and ethanol used. The functional unit was the switching cycle of optical properties of 50 x 50 x 3 mm OW CML 2001 was used as an established method to assess the switching cycle of optical properties of OW in

15 categories (acidification potential (AP), climate change potential (GWP), eutrophication potential (EP), freshwater aquatic ecotoxicity potential (FAETP), freshwater sediment ecotoxicity potential (FSETP), human toxicity potential (HTP), ionizing radiation (IR), land use (LU), malodorous air (MA), marine aquatic ecotoxicity potential (MAETP), marine sediment ecotoxicity potential (MSETP), photochemical oxidation (PCO), resources (R), stratospheric ozone depletion potential (ODP), and terrestrial ecotoxicity potential (TAETP)).

According to **Table S5 and Table S6**, the EIs of phenethyl alcohol and ethanol do not differ much during the optical performance cycle switching, with higher values of IR for phenethyl alcohol. The effects of AP, GWP and EP of phenethyl alcohol were approximately twice as high as those of ethanol. The MA, PCO, and OPD of phenethyl alcohol are only a quarter of those of ethanol. All other data were relatively similar. With the exception of MA, all other environmental impacts are relatively small, and proper packaging may significantly reduce the value of MA during use.

**Notes S1.** Numerical models for radiative cooling.
Cooling efficiency calculation
We begin by examining the radiative energy balance of the coolers under solar illumination. We take the cooler to be at temperature $T_{wood}$ and the ambient atmospheric temperature to be $T_{amb}$. The net cooling power density, defined as $P_{cool}$, is given by

$$P_{cool} = P_{rad}(T_{wood}) - P_{sun} - P_{atm}(T_{amb}) - P_{con}(T_{amb}, T_{wood}) \qquad (1)$$

Where $P_{rad}(T_{wood})$ is the power density of thermal radiation emitted by the cooling wood, $P_{sun}$ is the heating power density resulting from the absorption of solar irradiance, $P_{atm}(T_{amb})$ is the power density of the downward thermal radiation from the atmosphere, $P_{con}(T_{amb}, T_{wood})$ is the power density of thermal conduction and convection parasitically transferred to the cooler.

**Notes S2.** Numerical models for solar heating.
Heating-mode energy saving:
For heating energy saving, **Equation (2)** was used to analyze the device performance. $P_{heating}$ is the heating power of the device. $I$ is the solar radiation obtained. $\alpha$ is Transmission rate of heating materials.

$$P_{heating} = I\alpha \qquad (2)$$

The solar radiation obtained can be assessed by,

$$I = I_0 \sinh\left[P^m + \frac{1 - P^m}{2(1 - 1.4 \ln P)}\right] \qquad (3)$$

Where $I_0$ is the solar constant, $P$ is the atmospheric transparency coefficient, $m$ is the air mass, the solar elevation angle $h$ is given by,
$$sinh = sin\varphi sin\delta + cos\varphi cos\delta cos\omega \quad (4)$$
Where $\varphi$ is the local latitude ($\varphi_{Harbin}$ =45.85°), $\delta$ is the sun declination angle ($\delta_{Winter\ solstice}$) =-23.45°, $\omega$ is the hour angle.

**Notes S3.** Calculation of temperature difference due to sample thickness.
The heat absorption of the sample $Q$ is given by,
$$Q = \lambda \frac{A}{h}(T_{up} - T_{bottom}) \quad (5)$$
Where $\lambda$ is the thermal conductivity (The data in the text are obtained by testing), $T_{bottom}$ is the temperature of lower surface, $A$ is the sample area, $h$ is the thickness of sample.
1) Cooling section
The upper surface temperature $T_{OW-I}$ of OW-I can be obtained from the net power of OW-I.
According to **Figure S17**, the maximum value of solar spectral radiation power is 982 W/m². The average absorbance of the solar spectrum of wood is 0.115 (obtained by integration) and the emissivity is 0.93 (obtained by integration of the emissivity of the infrared spectrum)
According to Planck's law, Blackbody radiation law: The radiated heat:
$$Q_2 = \delta T_{OW-I}^4 * A * 0.93 \quad (6)$$

The heat generated by sunlight：
$$Q_1 = 982\ W/m^2 * A * 0.115 \quad (7)$$

The air heat radiation：
$$Q_3 = P_{atm}(T_{amb}) * A * 0.93 \quad (8)$$
The heat of OW-I
$$Q = Q_1 - Q_2 - h_{cc} * (T_{OW-I} - 306.5) * A + P_{atm}(T_{amb}) * A * 0.93 \quad (9)$$
The upper surface temperature of OW-I is $T_{OW-I}$ and bottom temperature is $T_{bottom\ OW-I}$ = 27.5°C= 300.8 K (**Figure 4c**), area $A$= 0.03*0.03 m², thermal conductivity $\lambda_{OW-I}$ = 0.1178 W m⁻¹ K⁻¹, $h_{OW-I}$ = 0.003 m, at 30 °C, $P_{atm}(T_{amb})$ =349.22 W/m²
$$Q_1 - Q_2 - h_{cc} * (T_{OW-I} - 306.5) * A + P_{atm}(T_{amb}) * A * 0.92 = A\lambda(T_{ow-I} - T_{bottom\ ow-I})/h \quad (10)$$
When $h_{cc}$=6 W m⁻² K⁻¹, $h_{cc}$ is convective heat transfer coefficient;
The calculated temperature of the upper surface of OW $T_{ow-I}$=296.0 K
The temperature difference between OW-I and air is: $\Delta T_{OW-I}$=306.5 K -296.0 K =10.5 °C
The temperature difference of OW-I by theoretical calculation is 7.5 °C, which is similar to the temperature difference by calculation the upper surface of OW-I ($\Delta T_{OW-I}$ =10.5 °C).
2) Photothermal section
The upper surface temperature $T_{NW}$ of NW can be obtained from the absorbed heat power of NW.
According to Planck's law, Blackbody radiation law: The radiated heat:
$$Q_2 = \delta T_{OW-I}^4 * A * 0.93 \quad (11)$$
According to Supplementary Figure 16, the maximum value of solar spectral radiation power is 982 W/m².

The heat generated by sunlight:

$$Q_1 = 982\ W/m^2 * A * 0.762 \tag{12}$$

The heat of OW-I

$$Q_1 - Q_2 - h_{cc} * (T_1 - 298.15) * A + P_{atm}(T_{amb}) * A * 0.93 \tag{13}$$

$T_{bottom\ NW}$= 51.9 °C = 325.1 K (**Figure 4c**), $A$= 0.03*0.03 m², $\lambda_{NW}$= 0.1363 W m⁻¹ K⁻¹. By consulting the spectra (atmospheric spectral data of the United States), at 30 °C, the integrating sphere calculates $P_{atm}\ (T_{amb})$ =349.22 W/m², thickness $h$ = 0.003 m

$$Q_1 - Q_2 - h_{cc} * (T_{NW} - 298.15) * A + 349.22 * A * 0.92 = A\lambda(T_{NW} - T_{bottom\ NW})/h \tag{14}$$

When $h_{cc}$=6 W m⁻² K⁻¹, $h_{cc}$ is convective heat transfer coefficient;
Form $T_{NW}$ =331.2 K
Temperature difference between NW and air is: $\Delta T_{NW}$ = 331.2 -306.5=24.7 °C
The temperature $\Delta T_{NW}$ and $\Delta T_{OW-I}$ on the upper surface of the sample were calculated to be consistent with the data in **Figure S13**.
The temperature difference of NW by theoretical calculation is 24.8 °C, which is basically the same as temperature difference by calculation the upper surface of NW ($\Delta T_{NW}$ =24.7 °C)

**Notes S4.** Comparison of the fabricated cost of optical wood and the current methods reported in the literature.
**This work (All prices are from the official Sigma-Aldrich website.)**
Price per unit:
Poplar 0.0956 Yuan cm⁻³ = 0.01488 dollars cm⁻³
NaClO₂ 0.177 dollars g⁻¹(1kg-177 dollars)
CH₃COOH 0.0428 dollars g⁻¹(2.5 kg-107 dollars)
CH₃COONa 0.0664 dollars g⁻¹(5 kg- 322 dollars)
Phenylethanol 0.3652 dollars g⁻¹ (25 kg- 913 dollars)
Ethanol 0.105 dollars/ mL (2 L- 210 dollars)
Dosage and Price:
Poplar 3 × 3 × 0.3 cm, 30 block, total 81 cm³, total 1.20528 dollars
NaClO₂ 12 g, total 2.124 dollars
CH₃COOH 2.4 g, total 0.10272 dollars
CH₃COONa 3.2 g, total 0.21248 dollars
Phenylethanol 200 g, total 73.04 dollars
Ethanol 200 mL, total 21 dollars
Total
97.68448 dollars
Unit cost
1.2060 dollars cm⁻³

| Reagent | Price per unit | Dosage | Total |
|---|---|---|---|
| Poplar | 0.01488 dollars cm$^{-3}$ | 81 cm$^3$ | 1.20528 dollars |
| NaClO$_2$ | 0.177 dollars g$^{-1}$ (1 kg- 177 dollars) | 12 g | 2.124 dollars |
| CH$_3$COOH | 0.0428 dollars g$^{-1}$ (2.5 kg- 107 dollars) | 2.4 g | 0.10272 dollars |
| CH$_3$COONa | 0.0664 dollars g$^{-1}$ (5 kg- 322 dollars) | 3.2 g | 0.21248 dollars |
| Phenylethanol | 0.3652 dollars g$^{-1}$ (25 kg- 913 dollars) | 200 g | 73.04 dollars |
| Ethanol | 0.105 dollars mL$^{-1}$ (2 L- 210 dollars) | 200 mL | 21 dollars |
| Total | — | — | 97.68448 dollars |
| **Unit cost** | — | — | **1.2060 dollars cm$^{-3}$** |

**Ref 1 Science 364, 760–763 (2019).**

Price per unit:

Poplar 0.0956 Yuan cm$^{-3}$ = 0.01488 dollars cm$^{-3}$

H$_2$O$_2$ 0.312 dollars mL$^{-1}$(500 mL- 156 dollars)

Ethanol 0.105 dollars mL$^{-1}$(2 L- 210 dollars)

1H, 1H, 2H, 2H-Perfluorooctyltriethoxysilane 18.64 dollars g$^{-1}$(25 g- 466 dollars)

Dosage and Price:

Poplar 3 × 3 × 0.3, 2.7 cm$^3$. 0.0956 × 10$^{-3}$ × 2.7=2.581 × 10$^{-2}$ dollars

H$_2$O$_2$ 50 × 0.312=15.6 dollars

Ethanol 70 × 0.105=7.35 dollars

1H, 1H, 2H, 2H-Perfluorooctyltriethoxysilane 0.3992 g × 18.64=7.441088 dollars

Total 2.5812 × 10$^{-2}$+15.6+7.35+7.441088=30.4169 dollars

Unit cost 30.4169/2.7=11.2655 dollars cm$^{-3}$

| Reagent | Price per unit | Dosage | Total |
|---|---|---|---|
| Poplar | 0.01488 dollars cm$^{-3}$ | 2.7 cm$^3$ | 2.581 × 10$^{-2}$ dollars |
| H$_2$O$_2$ | 0.312 dollars mL$^{-1}$ (500 mL- 156 dollars) | 50 mL | 15.6 dollars |
| Ethanol | 0.105 dollars mL$^{-1}$ (2 L- 210 dollars) | 70 mL | 7.35 dollars |
| 1H,1H,2H,2H-Perfluorooctyltriethoxysilane | 18.64 dollars g$^{-1}$ (25g- 466 dollars) | 0.3992 g | 7.4411 dollars |
| Total | — | — | 30.4169 dollars |
| **Unit cost** | — | — | **11.2655 dollars cm$^{-3}$** |

**Ref 2 Nat. Commun. 11, 1-9 (2020).**

Price per unit:

Polyimide 300 mm × 300 mm × 25 μm 2ea/ 469.00 dollars

Silver film (Ag, 99.99% PURE, 4.00" DIAMETER X 0.125") 6,995.00

copper film (Cu, 99.999% PURE, 4.00" DIAMETER X 0.125") 1,975.00
  Total (Add tax) 10,136.10=1569.0683 dollars
PDMS 3.72 dollars mL$^{-1}$(50 mL- 188 dollars)
Zn 0.392 dollars cm$^{-3}$(30*30 cm$^2$- 353 dollars)
ZnSO$_4$ 0.127 dollars g$^{-1}$(1 kg- 127 dollars)
CuSO$_4$ 0.342 dollars g$^{-1}$(500 g- 171 dollars)
Dosage and Price:
Polyimide 469/ (2 × 300 × 300) × 30 × 30=2.345 dollars
Zn 3 × 3=9 cm$^3$, 9 × 0.392=3.528 dollars
ZnSO$_4$ 4.03625 g × 0.127= 0.5126 dollars
CuSO$_4$ 19.1532 × 10$^{-3}$g × 0.342=6.5504 × 10$^{-3}$ dollars
PDMS 30 × 30 × 0.11= 99 mm$^3$ =0.099 mL, 0.099 × 3.72= 0.36828 dollars
Silver film & copper film
Price per unit 1569.0683/300=5.2302 dollars
Total
2.345+3.528+0.5126+6.5504 × 10$^{-3}$+0.36828+5.2302=11.9906 dollars
Unit cost
11.9906/27 × 1000=444.0974 dollars

| Reagent | Price per unit | Dosage | Total |
| --- | --- | --- | --- |
| Polyimide | 300 mm × 300 mm × 25 μm 2ea/ 469.00 dollars | 30 × 30 × 25 μm | 2.345 dollars |
| Silver film & copper film | 1569.0683 dollars | 0.003 | 5.2302 dollars |
| PDMS | 3.72 dollars mL$^{-1}$ (50 mL- 188 dollars) | 0.099 mL | 0.36828 dollars |
| Zn | 0.392 dollars cm$^{-3}$ (30 × 30 cm$^2$- 353 dollars) | 9 cm$^3$ | 3.528 dollars |
| ZnSO$_4$ | 0.127 dollars g$^{-1}$ (1 kg- 127 dollars) | 4.03625 g | 0.5126 dollars |
| CuSO4 | 0.342 dollars g$^{-1}$ (500 g- 171 dollars) | 19.1532 × 10$^{-3}$ g | 6.5504 × 10$^{-3}$ dollars |
| Total | — | — | 11.9906 dollars |
| **Unit cost** | — | — | **444.0974 dollars** |

**Ref 3 Joule 3, 3088-3099 (2019).**
Price per unit:
P(VdF-HFP) 0.543 dollars g$^{-1}$(100 g- 54.3 dollars)
Acetone 0.0706 dollars mL$^{-1}$ = 0.08938 dollars g$^{-1}$(1L- 70.6 dollars 1:8:1)
Isopropanol 0.123 dollars mL$^{-1}$ (500 mL- 61.5 dollars)
Dosage and Price:
Film 30 × 30 × 0.03 mm, total 27 mm$^3$ = 27 × 10$^{-3}$ mL, Solution volume 27 × 10$^{-2}$ mL.
Acetone 27 × 10$^{-2}$ × 0.7899 = 21.3273 × 10$^{-2}$ g, 1.9062 × 10$^{-2}$ dollars

P(VdF-HFP) 21.3273 × 10$^{-2}$/8 × 0.543=14.476 × 10$^{-3}$ dollars

Isopropanol

Cavity volume 30 × 30 × 2=1800 mm$^3$=1.8 mL, 1.8 × 0.123= 0.2214 dollars

Total 0.2214+1.4476 × 10$^{-2}$+1.9062 × 10$^{-2}$ = 0.254938 dollars

Unit cost 0.254938/27 × 1000 = 9.4421 dollars cm$^{-3}$

| Reagent | Price per unit | Dosage | Total |
| --- | --- | --- | --- |
| P(VdF-HFP) | 0.543 dollars g$^{-1}$ (100 g- 54.3 dollars) | 2.6659 × 10$^{-2}$ g | 1.4476 × 10$^{-2}$ dollars |
| Acetone | 0.08938 dollars g$^{-1}$ (1L- 70.6 dollars) | 0.2132 g | 1.9062 × 10$^{-2}$ dollars |
| Isopropanol | 0.123 dollars mL$^{-1}$ (500 mL- 61.5 dollars) | 1.8 mL | 0.2214 dollars |
| Total | — | — | 0.254938 dollars |
| **Unit cost** | — | — | **9.4421 dollars cm$^{-3}$** |

**Ref 4 Advanced Functional Materials, 2203582 (2022).**

Price per unit:

PVDF 3.11 dollars g$^{-1}$ (100 g- 311.01 dollars)

Cellulose acetate 0.4765 dollars g$^{-1}$ (500 g- 238.28 dollars)

Acetone 0.0706 dollars mL$^{-1}$ (1L- 70.6 dollars)

NMP 0.529 dollars mL$^{-1}$ (500 mL- 264.60 dollars)

PEG-400 2.5857 dollars g$^{-1}$ (1000 g- 258.75 dollars)

Ethylene glycol 0.118 dollars mL$^{-1}$ (1000 mL- 118.45 dollars)

Dosage and Price:

Film 10+2.5 =12.5 mL = 12.5 cm$^3$ (Calculated as half of the volume left after solvent removal.)

Acetone 10 × 0.0706 = 0.706 dollars

PVDF 2 × 3.11= 6.22 dollars

Cellulose acetate 1 × 0.4765= 0.4765 dollars

NMP 2.5 × 0.529 =1.3225 dollars

PEG-400 2 × 2.5857 = 5.1714 dollars

Ethylene glycol

Cavity volume 12.5 × 0.118 = 1.475 dollars

Total 0.706+6.22+0.4765+1.3225+5.1714+1.475 = 15.3714 dollars

Unit cost 15.3714 × 2 /12.5 = 2.4594 dollars cm$^{-3}$

| Reagent | Price per unit | Dosage | Total |
| --- | --- | --- | --- |
| PVDF | 3.11 dollars g$^{-1}$ (100 g- 311.01 dollars) | 2 g | 6.22 dollars |
| Cellulose acetate | 0.4765 dollars g$^{-1}$ (500 g- 238.28 dollars) | 1 g | 0.4765 dollars |
| Acetone | 0.0706 dollars mL$^{-1}$ (1L- 70.6 dollars) | 10 mL | 0.706 dollars |
| NMP | 0.529 dollars mL$^{-1}$ (500 mL- 264.60 dollars) | 2.5 mL | 1.3225 dollars |
| PEG-400 | 2.5857 dollars g$^{-1}$ (1000 g- 258.75 dollars) | 2 g | 5.1714 dollars |
| Ethylene glycol | 0.118 dollars mL$^{-1}$ (1000 mL- 118.45 dollars) | 12.5 mL | 1.475 dollars |
| Total | — | — | 15.3714 dollars |
| **Unit cost** | — | — | **2.4594 dollars cm$^{-3}$** |

**Ref 5 Advanced Functional Materials, 2208144 (2022).**

Price per unit:

SiO$_2$ 500 nm -3 μm 1.3277 dollars g$^{-1}$ (100 g- 132.77 dollars)

Ethanol 0.105 dollars mL$^{-1}$ (2 L- 210 dollars)

Carbon tetrachloride 1.3725 dollars mL$^{-1}$ (500 mL- 686.26 dollars)

Dosage and Price:

Film 70 × 70 × 40 × 10$^{-3}$ =196 mm$^3$ (Thickness calculated by SEM)

Ethanol 4 × 0.105 = 0.42 dollars

SiO$_2$ 0.2 × 1.3277= 0.2655 dollars

Carbon tetrachloride

Cavity volume 196 × 10$^{-3}$ × 2 × 1.3725= 0.53802 dollars

Total 0.42 + 0.2655 + 0.53802 = 1.2235 dollars

Unit cost 1.2235 × 10$^3$ /196 = 6.2423 dollars cm$^{-3}$

| Reagent | Price per unit | Dosage | Total |
| --- | --- | --- | --- |
| SiO$_2$ | 500 nm -3 μm 1.3277 dollars g$^{-1}$ (100 g- 132.77 dollars) | 0.2 g | 0.2655 dollars |
| Ethanol | 0.105 dollars mL$^{-1}$ (2 L- 210 dollars) | 4 mL | 0.42 dollars |
| Carbon tetrachloride | 1.3725 dollars mL$^{-1}$ (500 mL- 686.26 dollars) | 0.392 mL | 0.53802 dollars |
| Total | — | — | 1.2235 dollars |
| **Unit cost** | — | — | **6.2423 dollars cm$^{-3}$** |

**Ref 6 J. Mater. Chem. A, 10, 11092-11100 (2022).**

Price per unit:

N-isopropylacrylamine 5.94 dollars g$^{-1}$ (50 g- 297.03 dollars)

N, N'-Methylenebis (acrylamide) 0.5056 dollars g$^{-1}$(500 g- 252.81 dollars)

PVA 0.4699 dollars g$^{-1}$ (1 kg- 469.85 dollars)

2-Hydroxy-2-methylpropiophenone 14.58 dollars g$^{-1}$(5 g- 72.87 dollars)

PFOTS 16.28 dollars g$^{-1}$ =16.28/1.638 =9.94 dollars mL$^{-1}$(10 g- 162.75 dollars)

Dosage and Price:

Film 50 mL

N-isopropylacrylamine 5.93 × 5.94 = 35.28 dollars

N, N'-Methylenebis (acrylamide) 0.11 × 0.5056= 0.0556 dollars

PVA 0.66 × 0.4699=0.3101 dollars

2-Hydroxy-2-methylpropiophenone 0.079 ×14.58= 1.1518 dollars

PFOTS 5 × 9.94= 49.7 dollars

Total 35.28 + 0.0556 + 0.3101+1.1517 + 49.7 = 86.4974 dollars

Unit cost 86.4974 /50 = 1.7299 dollars cm$^{-3}$

| Reagent | Price per unit | Dosage | Total |
| --- | --- | --- | --- |
| N-isopropylacrylamine | 5.94 dollars g$^{-1}$ (50 g- 297.03 dollars) | 5.93 g | 35.28 dollars |
| N, N'-Methylenebis (acrylamide) | 0.5056 dollars g$^{-1}$ (500 g- 252.81 dollars) | 0.11 g | 0.0556 dollars |
| PVA | dollars | 0.66 g | 0.3101 dollars |
| 2-Hydroxy-2-methylpropiophenone | 14.58 dollars g$^{-1}$ (5 g- 72.87 dollars) | 0.079 g | 1.1518 dollars |
| PFOTS | 16.28 dollars g$^{-1}$ =16.28/1.638 =9.94 dollars mL$^{-1}$ (10 g- 162.75 dollars) | 5 mL | 49.7 dollars |
| Total | — | — | 86.4974 dollars |
| **Unit cost** | — | — | **1.7299 dollars cm$^{-3}$** |

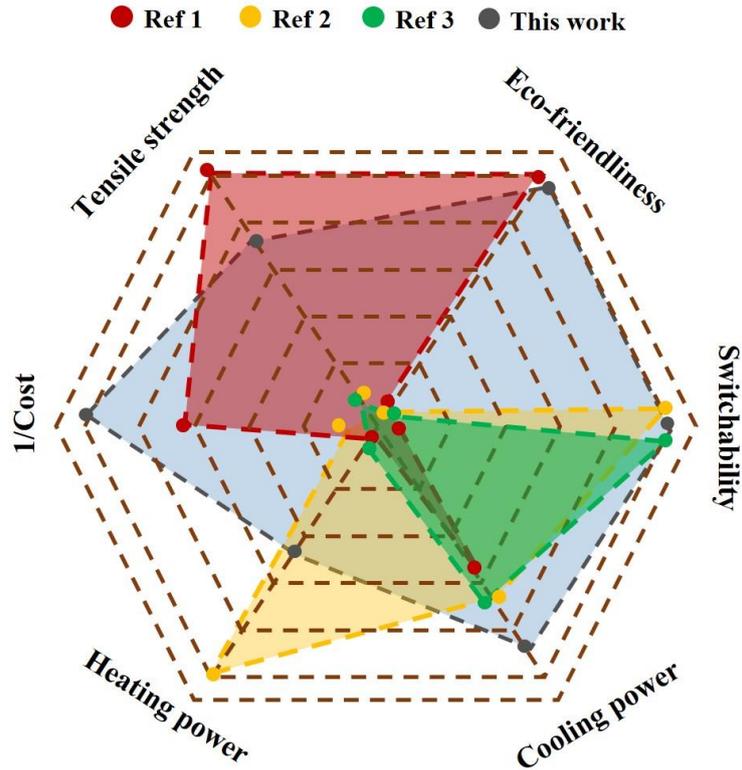

**Figure S22.** The radar map composed of optical wood and current methods reported in the literature.

**Table S7** Comparison of six character of optical wood and current methods reported in the literature.

|  | Ref.1 | Ref.2 | Ref.3 | Ref.4 | Ref.5 | Ref.6 | This work |
|---|---|---|---|---|---|---|---|
| 1/Cost | 1/11.27 | 1/444.10 | 1/9.44 | **1/2.46** | 1/6.24 | **1/ 1.73** | **1/1.21** |
| Tensile strength | **404.3** | 0 | 0 | 5.4 | 0 | 0 | 22.3 |
| Cooling power | 63 | 71.6 | ~71 | 43.4 | **85.1** | 47.0 | **81.4** |
| Natural | **100** | 0 | 0 | 0 | 0 | 0 | **100** |
| Heating power | 0 | 643.4 | Not stated | **744** | Not stated | 197 | 229.5 |
| Switchable | 0 | **100** | **100** | **100** | **100** | **100** | **100** |

**References**
1. Li T, Zhai Y, He S, Gan W, Wei Z, Heidarinejad M, Dalgo D, Mi R, Zhao X, Song J, Dai J, Chen C, Aili A, Vellore A, Martini A, Yang R, Srebric J, Yin X, Hu L, Science 2019; 364: 760-763.
2. Li X, Sun B, Sui C, Nandi A, Fang H, Peng Y, Tan G, Hsu PC, Nat. Commun. 2020; 11: 1-9.
3. Mandal J, Jia M, Overvig A, Fu Y, Che E, Yu N, Yang Y, Joule 2019; 3: 3088-3099.
4. Fei J, Han D, Ge J, Wang X, Koh SW, Gao S, Sun Z, Wan MP, Ng BF, Cai L, Li H, Adv. Funct. Mater. 2022; 2203582.